\documentclass[a4paper,16pt]{article}
\usepackage[noadjust]{cite}
\usepackage{amsfonts}
\usepackage{graphicx,epstopdf}
\usepackage{caption}
\usepackage{subcaption}
\usepackage{siunitx}
\usepackage[colorlinks=true,linkcolor=blue,anchorcolor=blue,urlcolor=blue,citecolor=blue]{hyperref}
\DeclareGraphicsExtensions{.png,.eps,.pdf}
\usepackage{amsmath}
\usepackage{amsfonts}
\usepackage{latexsym}
\usepackage{amssymb}
\usepackage{verbatim}
\usepackage{multirow}
\usepackage{array}
\usepackage[percent]{overpic}
\usepackage{xcolor}
\usepackage{geometry}
\newcolumntype{P}[1]{>{\centering\arraybackslash}p{#1}}
\newcolumntype{M}[1]{>{\centering\arraybackslash}m{#1}}
\usepackage{colortbl}
\usepackage{booktabs}
\usepackage{eurosym}
\usepackage{pgfplotstable}

\usepackage{pgfplots}
\pgfplotsset{compat=newest}

\usepackage{tikz}
\usetikzlibrary{patterns}
\usetikzlibrary{shapes,arrows,calc}

\title{Propeller Tip Vortex Mitigation By Roughness Application}

\usepackage{authblk}
\author[1]{Abolfazl Asnaghi}
\author[2]{Urban Svennberg}
\author[2]{Robert Gustafsson}
\author[1,a]{Rickard E. Bensow}
\affil[1]{\small Department of Mechanics and Maritime Sciences, Chalmers University of Technology, 41296 Gothenburg, Sweden}
\affil[2]{Kongsberg Hydrodynamic Research Centre, Kongsberg Maritime Sweden AB, 68195 Kristinehamn, Sweden}
\affil[a] {Author to whom correspondence should be addressed: rickard.bensow@chalmers.se}

\begin{document}

\maketitle
\section{Abstract}
In this study, the application of surface roughness on model and full scale marine propellers in order to mitigate tip vortex cavitation is evaluated. To model the turbulence, SST $k-\omega$ model along with a curvature correction is employed to simulate the flow on an appropriate grid resolution for tip vortex propagation, at least 32 cells per vortex diameter according to our previous guidelines. The effect of roughness is modelled by modified wall functions.

\noindent
The analysis focuses on two types of vortices appearing on marine propellers: tip vortices developing in lower advance ratio numbers and leading edge tip vortices developing in higher advance ratio numbers. It is shown that as the origin and formation of these two types of vortices differ, different roughness patterns are needed to mitigate them with respect to performance degradation of propeller performance. Our findings clarify that the combination of having roughness on the blade tip and a limited area on the leading edge is the optimum roughness pattern where a reasonable balance between tip vortex cavitation mitigation and performance degradation can be achieved. This pattern in model scale condition leads to an average TVC mitigation of 37\% with an average performance degradation of 1.8\% while in  full scale condition an average TVC mitigation of 22\% and performance degradation of 1.4\% are obtained.

\noindent
Keywords: Tip vortex, Mitigation, Roughness, Propeller, CFD.

 
\newpage 
\section{Introduction}
\noindent
A hydrodynamically optimum propeller design usually does not have an optimum hydroacoustic performance as their design restrictions are contradictory ~\cite{Savas2016}. This has even further importance for low-noise propellers as their operating profile requires very low radiated noise emissions mostly generated by cavitation ~\cite{Kuiper1981}. Tip vortex cavitation (TVC) is usually the first type of cavitation that appears on a propeller, and consequently plays a key role in initiating an overall increasing sound pressure level, and determining the underwater radiated noise ~\cite{Arndt20150025,BosschersPhDThesis}. The radiated noise is of big importance because it can disturb marine wildlife and reduces comfort of people on board ship. Therefore, TVC is considered as the main cavitation characteristics to control in the in the design procedure. 

\noindent
The blade load distribution is a decisive parameter in the tip vortex cavitation formation ~\cite{Kuiper2001}. With a highly loaded tip, vorticity is generated at the trailing edge of a propeller blade, resulting in a stable trailing vortex formation. With reduced tip loading, separation still occurs close to the tip, while the trailing vortex is much weaker, leading to a local tip vortex formation typical for a standard ship propeller. An unloaded tip design forces the loading towards inner radii and at these inner radii, leading edge separation, and therefore a leading edge vortex, may be formed and the trailing vortex system becomes more distributed \cite{Lu2014}. In this condition, effects of non-uniform flow field ~\cite{jmse4040070,Korkut200229}, and blade surface roughness ~\cite{Matthieu2015, Felli20111385} should also be considered.

\noindent
The start of cavitation in the tip vortex, TVC inception (TVCI), determines a break point where nuisance suddenly increase. Cavitation will occur in the core of a tip vortex only if a nucleus has enough time to reach the core and then trigger cavitation ~\cite{Boulon1997752}. It is well reported that cavitating behaviour of a tip vortex depends on the nuclei radius, its initial location, vortex circulation, and vortex velocity ~\cite{Arndt1992,Ligneul1993504,Briancon97,Hsu1991504}. Depending on the water quality, TVCI can be either a sudden appearance of a continuous cavity, or an intermittent appearance of an elongated bubble extending axially over a relatively small portion of the tip vortex ~\cite{Higuchi1989495,Maines1997271}. At the intermittent step, formation and collapse of elongated bubbles increases noise intensity significantly compared with the fully developed TVC ~\cite{Pennings2015S}. Looking at vortex singing, the tones are very different under different flow conditions and water quality where the vortex singing can only exist in the transferring process between strong tip vortex cavitation and weak tip vortex cavitation  ~\cite{Peng2015, Peng2017}.

\noindent
Traditional potential flow propeller design tools, in connection with designer experience, are able to provide optimal geometries in terms of efficiency, and to some extent capture the effects of on-blade cavitation, but they are not suited for the assessment of negative aspects of cavitating tip vortices. Apart from redesigning a propeller to redistribute load and consequently changing TVC properties ~\cite{Shin2015}, a few approaches has been proposed to suppress TVC, classified as active control and passive control methods. In active control approaches, the tip vortex flow is altered by injection of a solution, e.g. air ~\cite{Rivetti2014}, polymer ~\cite{Fruman1989211,Yakushiji2009}, or water ~\cite{Chahine1993497,Chang2011}, into the tip vortex region. In passive control methods, the boundary layer and momentum distribution around the blade tip are altered aiming to weaken the tip vortex and its nuclei capture capability. Inclusion of an extra geometry on the blade tip ~\cite{Gim201328,Park20141,Brown2015}, drilling holes on the blade ~\cite{AktasSMP2019}, and roughening the blade surface ~\cite{Johnsson91,Kruger2016110,bhl102158,asnaghiSMP2019} are some examples of passive TVC mitigation, of which the latter is the concern of this study.

\noindent
Surface roughness affects the tip vortex roll-up as roughness elements promote transition to turbulence and growth of laminar boundary layers and thereby alter the near-wall flow structures. The vortical structures generated by the roughness elements interact with the main tip vortex and destabilize it. If size, pattern, and location of roughness elements are selected appropriately, the destabilization process leads to tip vortex breakdown, and consequently to TVC mitigation. This, however, increases the losses and leads to performance degradation \cite{Johnsson91,Kruger2016110,asnaghiSMP2019}. To minimize this degradation, one has to optimize the roughened area which itself demands a detailed knowledge on where and how the tip vortex is formed. For propellers  in behind conditions, this is even more complex as the tip vortex location and origin vary. 

\noindent
In the current study, different flow properties are analyzed to define effective areas in tip vortex formation and its roll-up on a propeller selected from a research series of highly skewed propellers having a low effective tip load which is typical for yachts and cruise ships. In our previous studies, numerical simulations of tip vortex flows around this propeller having smooth blades have been carried out and successfully compared with experimental measurements \cite{ASNAGHI2018197}. The aim of the present study is to provide further knowledge about the effects of the surface roughness on the TVC and the possibility of using roughness to delay the cavitation inception. 

\noindent
The turbulent flow field around the propellers is modelled by using the two-equation SST $k-\omega$ model of OpenFOAM on appropriate grid resolutions for tip vortex propagation, at least 32 cells per vortex diameter according to previous studies guidelines \cite{AsnaghiPhDThesis, ASNAGHI2019135}. To prevent overprediction of turbulent viscosity in highly swirling tip regions, $\eta_3$ curvature correction method is employed \cite{ArolaThesis,Arola2014}. The roughness is included in the simulations by employing two different approaches. In the first approach, rough wall functions are used to mimic the effects of roughness by modifying the turbulent properties in roughed areas \cite{TapiaMaster2009}. The second approach modifies the mesh topology by removing cells in roughed areas to create random roughness elements. See further comment below on the resolved geometry.

\noindent
To identify the areas where roughness has to be applied, three criteria based on the flow properties of the smooth propeller condition are employed. The Q-criterion is selected to identify the vortical structures and their interactions. The second criterion is the flow streamlines close to the blade surface helping to highlight from which areas on the blade the vortex momentum is provided. The third criterion is the pressure coefficient used to distinguish the areas where the cavitation incepts inside the tip vortex. These criteria are used to find a good roughness pattern that can lead to a proper tip or leading edge vortices mitigation with reasonable performance degradation.

\noindent
The strategy used to find the suitable roughness pattern in model scale condition is extended to find the roughness pattern in full scale conditions. This has been done by considering the flow properties of the smooth propeller in the full scale condition as well. Then, the best roughness pattern application is evaluated at three operating conditions, i.e. J=0.82, 0.93 and 1.26. Contradictory to the model scale analysis where a rough wall function is employed to incorporate the effects of roughness, in the full scale analysis the roughness elements are included as a part of computational domain. This gives the possibility of resolving the flow around the roughness elements, and also allows to use smaller y$^+$ for the blade surfaces.

\noindent 
The results contain the performance and cavitation inception charts of the propeller in model and full scale conditions. Roughness application on different blade areas are examined at different operating conditions, and their impact on TVC mitigation and performance degaradtion is reported. It is investigated how having different roughness patterns alter vortical structures on the blade and in the tip vortex region. The roughness area is optimized by simultaneous consideration of the tip vortex mitigation, performance degradation and their compromise. The performance and TVC mitigation of the propeller having optimum roughness pattern are discussed.      	
\section{Governing Equations}
 \noindent
The OpenFOAM package, used in this study for numerical simulation, is an open source code written in C++ to model and simulate fluid dynamics and continuum mechanics \cite{OpenFOAM}. The incompressible conservation of mass and momentum equations are solved using the PIMPLE algorithm, a merge of the SIMPLE and PISO algorithms. The solver has been used and validated for tip vortex analysis in marine applications; see \cite{AsnaghiPhDThesis,ASNAGHI2020106703} for more details on the modelling and the numerical setup used in OpenFOAM.

\subsection{Turbulence model}
The turbulence is modelled by employing the SST $k-\omega$ model along with a curvature correction model \cite{ASNAGHI2019135}. In the selected model, the production term of the $\omega$ equation is multiplied by $F_{rc}$,
\begin{equation}
\label{Eq.etha3Based}
	F_{rc}= 1 + \alpha_1 \mid\eta_3\mid + 3 \alpha_1 \eta_3 ,
\end{equation}
where $\alpha_1=-0.2$ and $C_r=2.0$. In this equation, $\eta_3$ is a velocity gradient invariants defined through the non-dimensional strain rate and rotational rate tensors \cite {ArolaThesis},
\begin{equation}
\label{Eq.etaParameters}
	\eta_1 =\bar S_{ij}^* \bar S_{ij}^*, \hspace{0.3cm} \eta_2 =\bar  \Omega_{ij}^* \bar  \Omega_{ij}^*, \hspace{0.3cm} \eta_3 = \eta_1 -\eta_2,
\end{equation}
\begin{equation}
\label{Eq.sijOmegaijNon}
	\bar S_{ij}^{*}=\tau \bar S_{ij}, \quad \bar \Omega_{ij}^{*}=\tau \bar \Omega_{ij}^{mod},
\end{equation}
\noindent
where the strain rate and rotational rate tensors are defined by,
\begin{equation}
\label{Eq.sij}
	\bar S_{ij}=\frac{1}{2}(\frac{\partial \bar u_{i}}{\partial x_{j}}+\frac{\partial \bar u_{j}}{\partial x_{i}}), \quad \bar \Omega_{ij}=\frac{1}{2}(\frac{\partial \bar  u_{i}}{\partial x_{j}}-\frac{\partial \bar  u_{j}}{\partial x_{i}}).
\end{equation}

\noindent
As can be seen, $\eta_1$ represents the non-dimensional strain rate magnitude, $\eta_2$ represents the non-dimensional vorticity magnitude, and $\eta_3$ is a linear combination of these two velocity-gradient invariants. The turbulent time scale used to non-dimensionalize these tensors is calculated by \cite {ArolaThesis,starCCM,Arola2014},
\begin{equation}
\label{Eq.turbTimeScaleDef}
	\tau = \text{max}(\tau_1,\tau_3),\hspace{0.3cm}\tau_1 = \frac {1}{\beta^*\omega},\hspace{0.3cm}\tau_2= 6\sqrt {\frac {\nu}{\beta^*k\omega }},\hspace{0.3cm}\tau_3 =(\tau_1^n \tau_2)^{\frac{1}{n+1}},
\end{equation}
where $n=1.625$. The time scale is limited in order to have a correct near-wall asymptotic behaviour. 

\noindent
The modified rotational rate tensor incorporating the streamline curvature and frame rotation is,
\begin{equation}
\label{Eq.omegaijMod}
	\bar \Omega_{ij}^{mod}=\bar \Omega_{ij}+ \Omega^F_{ij} + (C_r-1)W_{ij}^{A},
\end{equation}
where $C_r$ is the constant of the equation and depends on the CC model \cite {ArolaThesis}. This coefficient takes a value of 2 for bifurcation approaches. Here, $\Omega^F_{ij}$ represents the frame rotational tensor calculated from $\Omega^F_{ij}=-\epsilon_{ijk}\Omega^F_k$ where $\Omega^F_k$ is the angular frame velocity about the $x_k$-axis. The $W_{ij}^{A}$ tensor which contains the effects of curvature corrections in the rotational rate tensor is defined by \cite {Wallin2002721},
\begin{equation}
\label{Eq.omegaStarRo}
	W_{ij}^{A} = -\epsilon_{ijk} B_{km} \bar  S_{pr} \frac {D \bar  S_{rq}}{Dt} \epsilon_{pqm} ,
\end{equation}
\begin{equation}
\label{Eq.bKM}
	B_{km}=\frac{II_S^2 \delta_{km}+12 III_S \bar S_{km} + 6 II_S \bar S_{kl}  \bar S_{lm}}{2 II_S^3-12III_S^2},
\end{equation}
\begin{equation}
\label{Eq.IISIIIS}
	II_S =\bar S_{kl} \bar S_{lk},\hspace{0.4cm} III_S =\bar S_{kl} \bar S_{lm} \bar S_{mk},
\end{equation}
where $\frac {D \bar S_{rq}}{Dt}$ is the material derivative of the strain rate tensor. Please refer to \cite{AsnaghiPhDThesis,ArolaThesis} for further information.

\subsection{Roughness modelling}
 \noindent
The flow around the roughness elements can be either resolved or modelled. To resolve the flow, roughness geometries have to be included into the computational domain which leads to having finer cell resolution around them compared to the rest of the domain, and consequently demand for higher computational resources. Modelling roughness elements requires much lower number of computational cells but as it involves simplification of the roughness geometry, the flow physics may not be correctly modelled.

\noindent
In the current study for modelling roughness elements, the wall function developed by Tapia (2009) for the inner region of the turbulent boundary layer or the log-law region (e.g. $11 \leq y^+$ in OpenFOAM wall functions) is used,  
\begin{equation}
\label{Eq.uPlusEquation}
u^{+}=\frac {1}{\kappa}\text{ln}(E y^{+})-\Delta B,
\end{equation}
with the von Karman constant $\kappa=0.41$, the constant E=9.8, the dimensionless wall distance $y^{+}=u_\tau y /\nu$, and the velocity shift correction $\Delta B$ due to the roughness elements. In this model, the nondimensional roughness height is presented by $K_s^{+}=u_\tau K_s /\nu$ where $K_s$ is the roughness height, $u_\tau = \sqrt {\tau_w / \rho}$ is the shear velocity, and $\tau_w$ is the wall shear stress. As this model only affects the viscosity of the first cell adjacent to the wall, the height of the roughness elements should be smaller than the height of the cells wall normal distance, i.e. $\text{K}_\text{s}^{+}  \leq \text{y}_\text{w}^{+}$. Otherwise, the part of roughness elements that locates outside the adjacent cells will not be included in the modelling.

\noindent
The flow regime over a rough surface depends on how roughness elements interact with different parts of the boundary layer. If the roughness elements are embedded in the viscous sublayer, the friction drag is not affected by the roughness and the flow regime is smooth. In a smooth regime represented by $\text{K}_\text{s}^{+} \leq 2.5$, the correction $\Delta B$ is set to zero and the wall function recalls the smooth wall function.

\noindent
In the case where the roughness element heights are much larger than the boundary layer thickness, the fully rough regime forms where the drag significantly increases. In such a condition, the pressure drag on the roughness elements dominates and the impact of roughness becomes independent of Reynolds number which means the viscous effect is no longer important. For a fully rough regime represented by $90 \leq \text{K}_\text{s}^{+}$, the $\Delta B$ correction is represented by,
\begin{equation}
\label{Eq.fullyRoughWall}
\Delta B = \frac{1}{\kappa}\text{ln}\bigg[1+\text{C}_\text{s}\text{K}_\text{s}^{+}\bigg].
\end{equation}

\noindent
The transition regime happens where both viscous and pressure forces on the roughness elements contribute to the wall skin friction. In this condition represented by $2.5<\text{K}_\text{s}^{+}<90$, the correction reads,
\begin{equation}
\label{Eq.transitionallyRoughWall}
\Delta B = \frac{1}{\kappa}\text{ln}\bigg[\frac{\text{K}_\text{s}^{+}-2.25}{87.75}+\text{C}_\text{s}\text{K}_\text{s}^{+}\bigg]\text{sin}\big(0.425[\text{ln}(\text{K}_\text{s}^{+})-0.811]\big).
\end{equation}

\noindent
In these equations, shape and form of roughness elements are incorporated into the modelling through the $C_\text{s}$ coefficient. However, there is no clear guideline to adjust this coefficient. It is suggested that it varies from 0.5 to 1 where $\text{C}_\text{s}$=0.5 corresponds to the uniformly distributed sand grain roughness. If the roughness elements deviate from the sand grains, the constant roughness should be adjusted by comparing the results with experimental data.   

\subsection{Parameters defining the flow properties} 
  \noindent
The hydrodynamic performance of a propeller is defined by using the non-dimensional thrust and torque coefficients, and the advance ratio,
\begin{equation}
\label{Eq.ThrustCoeff}
     K_T=\frac {T}{\rho n^2 D^4},\quad K_Q=\frac {Q}{\rho n^2 D^5},\quad J=\frac {V_A}{n D}.
\end{equation}
In these equations, $D$ is the propeller diameter, $n$ is the rotational speed of the propeller in rev/sec, $\rho$ is the fluid density, $T$ is the propeller thrust force, $Q$ is the propeller shaft torque, and $V_A$ is the mean inflow velocity towards the propeller plane.

\noindent
To identify vortical structures in the flow, the Q-criterion representing the local balance between shear strain and rotational tensor magnitudes, is employed ~\cite{Hunt1998, Kolar2007},
\begin{equation}
\label{Eq.QCriterion}
    Q=\frac{1}{2}(\bar \Omega_{ij}\bar \Omega_{ij} - \bar S_{ij}\bar S_{ij}).
\end{equation}

\noindent
To simplify the cavitation inception detection, the minimum pressure criterion is employed \cite{ASNAGHI2018197}. The criterion assumes that cavitation occurs as soon as the minimum pressure in the flow reaches the saturation pressure. Therefore, the cavitation inception point, $\sigma_i$, is determined from the pressure field of the wet flow as,
\begin{equation}
\label{Eq.minPressure}
\sigma_i=-C_{p,min}.
\end{equation}

\section{Case description}
The basic design of the propeller is from a research series of five-bladed highly skewed propellers having low effective tip load for vessels where it is very important to suppress and limit propeller-induced vibration and noise. The main, or first occuring source of noise, for this type of propellers, is cavitation in the tip region. In the previous studies conducted by the authors \cite{AsnaghiPhDThesis,ASNAGHI2020106703,ASNAGHI2018197}, the computational guideline to successfully model the tip vortex flow around this propeller having smooth surface within OpenFOAM was investigated. In this guideline, the turbulence modelling impact, minimum required spatial mesh resolution for modelling the tip vortex in the near field region and the numerical set up that can provide low numerical dissipation and high stability are discussed. Here, the same guideline for the computational domain and mesh specifications is employed.

\noindent
The computational meshes were generated by StarCCM+ and then converted into the OpenFOAM format. As the open water conditions was of interest, simulation of the flow around one blade and then using cyclic boundaries on the sides were possible. However, when the numerical analysis of the model scale propeller was conducted, there were some stability issues with the cyclic boundaries making it very cumbersome to converge. Therefore, the whole propeller is modelled and in order to keep the number of cells low enough, the tip vortex refinement was applied on one blade only. The utility used to convert StarCCM+ mesh to the OpenFOAM format has a limitation on the number of cells or faces which forced using the Trimmer mesher to create the model scale propeller mesh. For the full scale propeller, using the Trimmer mesher led to some bad quality cells, e.g. wrongly oriented faces or face non-orthogonality more than 85 degrees. This problem is believed to be related to the low numerical precision of reading or writing the points storing the locations of cells and faces. However, during the study it was not possible to find out where this low reading or writing precision occurs. At the same time as we were able to improve the stability of employing cyclic boundaries, the polyhedral mesher is used to create the mesh around the full scale blade.

\begin{figure}[h!]
        \centering
        \begin{subfigure}[b]{0.485\textwidth}
				\includegraphics[width=\textwidth]{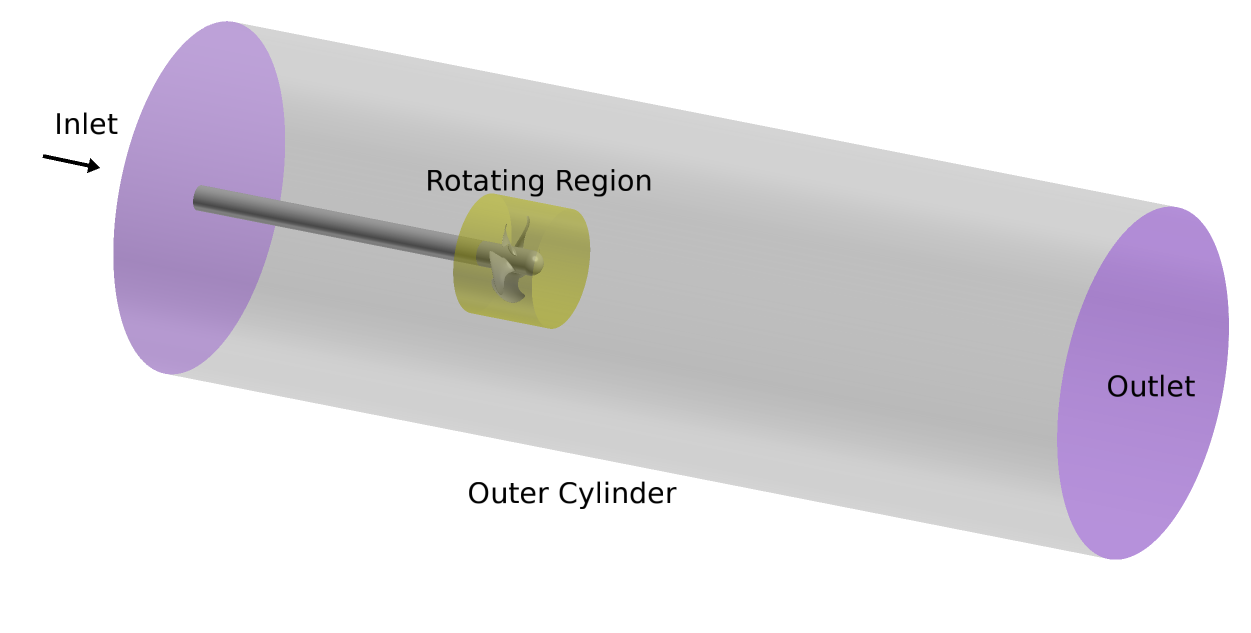}
			\caption{Computational domain}		
			\label{fig::starCCMmeshDisRRProepllerMSDomain}												
        \end{subfigure}
		\quad        
        \begin{subfigure}[b]{0.485\textwidth}
				\includegraphics[width=\textwidth]{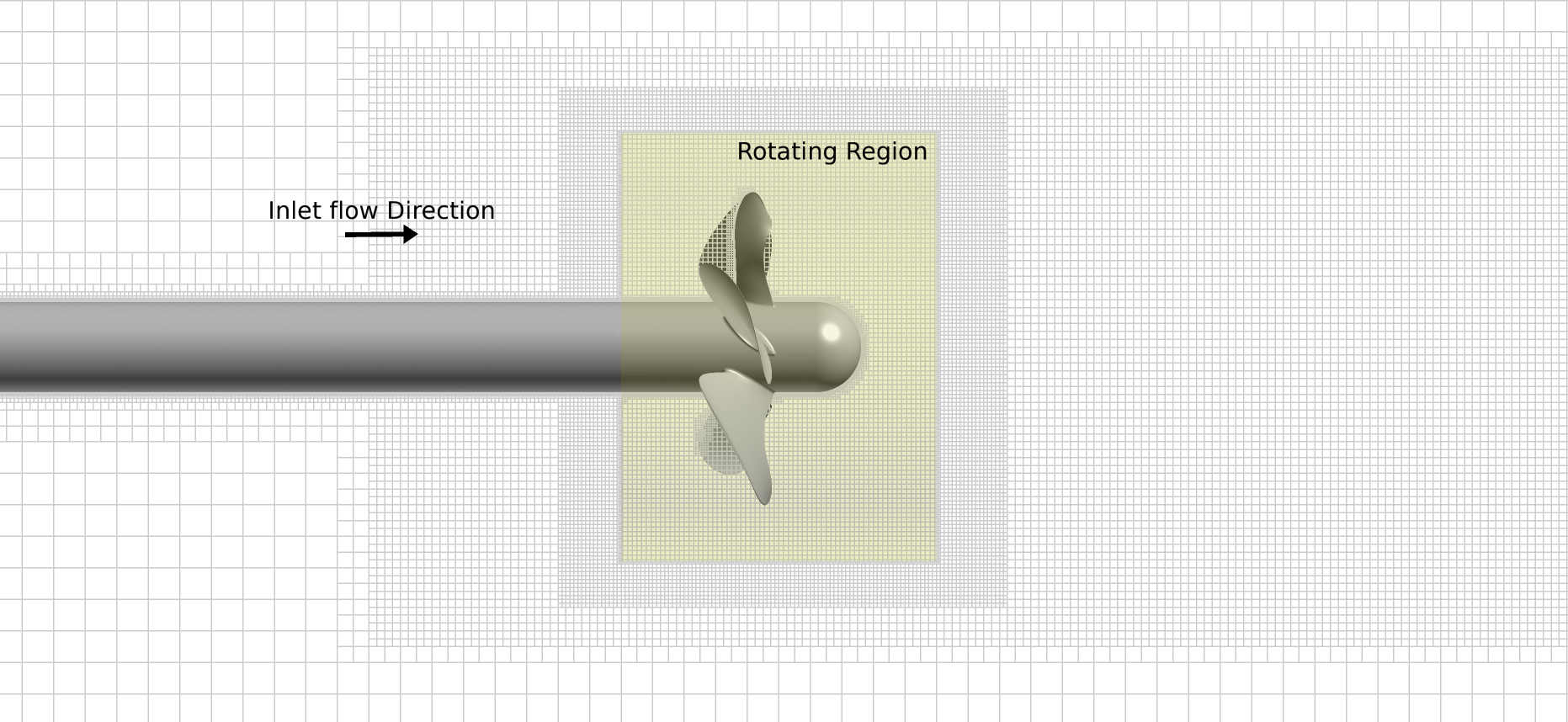}
			\caption{Streamwise resolution}		
			\label{fig::starCCMmeshDisRRProepllerDomain}												
        \end{subfigure}
		\quad
        \begin{subfigure}[b]{0.35\textwidth}
				\includegraphics[width=0.975\textwidth]{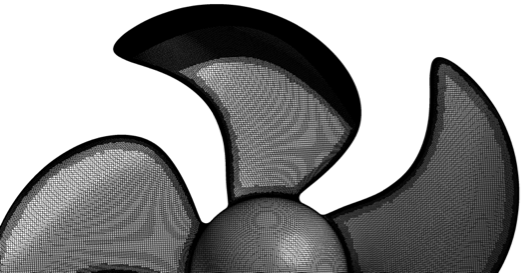}
			\caption{Blade surface resolution}
			\label{fig::starCCMmeshDisRRProepllerBlade}								
        \end{subfigure}
		\quad
        \begin{subfigure}[b]{0.35\textwidth}
				\includegraphics[width=0.975\textwidth]{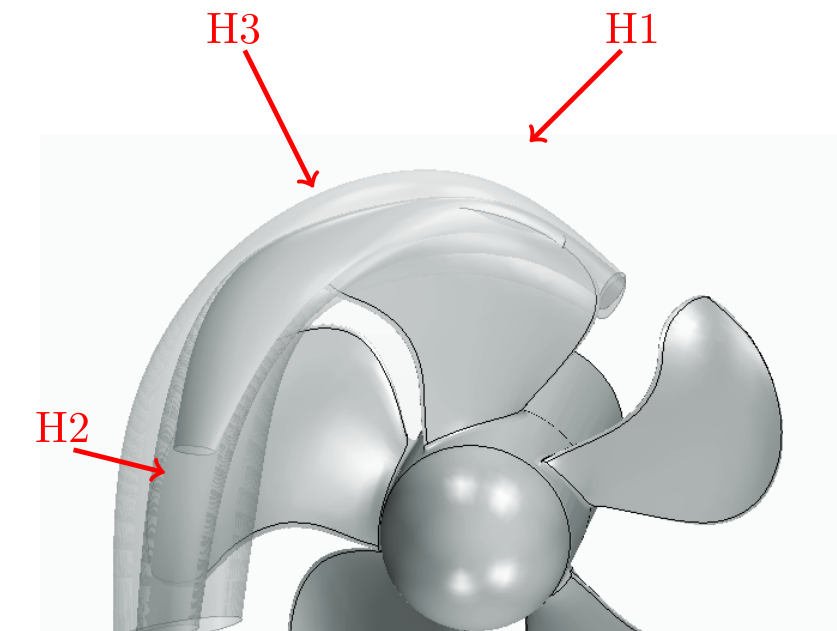}
		\caption{Helical tip refinement}												
        \end{subfigure}
                		               
		\caption{Mesh distribution of the model scale propeller.}
		\label{fig::starCCMmeshDisRRProepller}
\end{figure}

\noindent
The computational domain used for the model scale propeller is presented in Figure \ref{fig::starCCMmeshDisRRProepllerMSDomain}. The domain is simplified to a cylinder extending 4D upstream the propeller and 8D downstream of the propeller where D=0.2543 m is the diameter of the propeller. The simulations are conducted at a constant inlet velocity, a fixed pressure outlet boundary and the advance ratio of the propeller is then set by adjusting the rotational rate of the propeller. No-slip wall boundary condition is used for the propeller and the shaft. The outer cylinder boundary is set as a slip boundary to reduce the mesh resolution requirements far from the propeller. Then, in order to consider the blockage effects, the numerical results are compared against the experimental measurements which would provide the same velocity at the cavitation tunnel measurement point. In order to model the moving mesh (i.e. relative motion between the propeller and the external domain), the computational domain has been decomposed into two regions connected to each other through AMI (Arbitrary Mesh Interpolation) boundaries. While the outer region is stationary, the rotation of the region close to the propeller where all interesting flow phenomena occur has been handled by MRF. 

\noindent
Different refinement boxes are applied to provide finer resolutions around the rotating propeller region, Figure \ref{fig::starCCMmeshDisRRProepllerDomain}. The baseline mesh resolution on the blades gives ${x^+}$ and ${z^+} < 250$, with much finer resolutions at the leading edge and trailing edge of the blades due to the high geometry curvature. Here, ${x^+}$ and ${z^+}$ are non-dimensionalized resolutions in x and z directions calculated similar to $y^{+}$ term, i.e. $x^{+}=u_\tau \Delta x /\nu$ and $z^{+}=u_\tau \Delta z /\nu$. In these definitions, $\Delta x$ and $\Delta z$ are the cell resolution in x and z directions, respectively. As the surface resolution close to the tip is determined by the tip refinement boxes, even finer resolution is achieved at the blade tip and the leading edge. The prismatic layers of the refined blade consists of 20 layers having extrusion factor of 1.15 where the first cell wall normal resolution is set equal to $y^+ = 35$ for wall modelling simulations, and equal to $y^+ = 5$ for resolving the flow around the roughness elements. As mentioned, the tip vortex refinement is applied on one blade only where three helical shape refinement zones are defined based on the primary vortex trajectory. The refinement zones cover the tip of the blade, and therefore provide more refined grid resolutions on the tip of this blade, Figure \ref{fig::starCCMmeshDisRRProepllerBlade}. These helical refinement regions provide spatial resolutions as fine as 0.2 mm, 0.1 mm, and 0.05 mm in H1, H2, and H3 regions for the model scale propeller mesh, respectively.

\begin{figure}[h!]
        \centering
        \begin{subfigure}[b]{0.47\textwidth}
				\includegraphics[width=\textwidth]{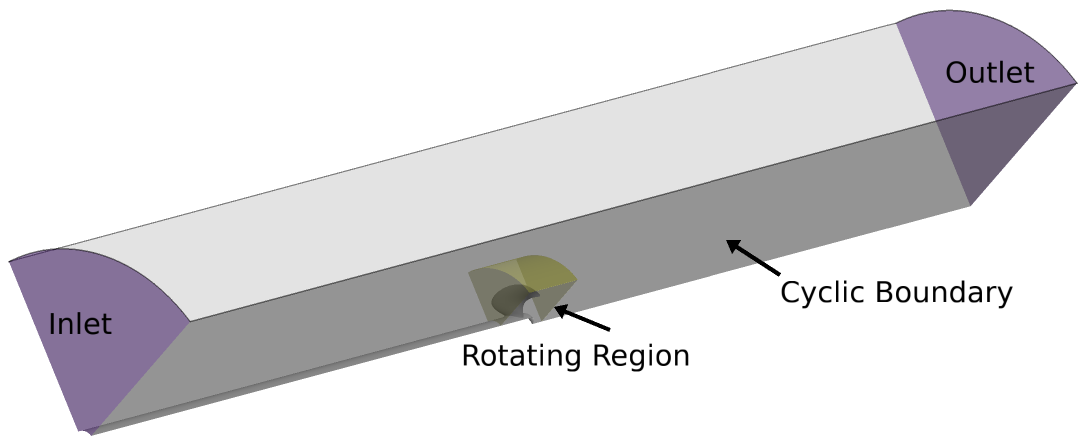}
			\caption{Computational domain}		
			\label{fig::starCCMmeshDisRRProepllerFSDomain}												
        \end{subfigure}
		\quad
        \begin{subfigure}[b]{0.47\textwidth}
				\includegraphics[width=\textwidth]{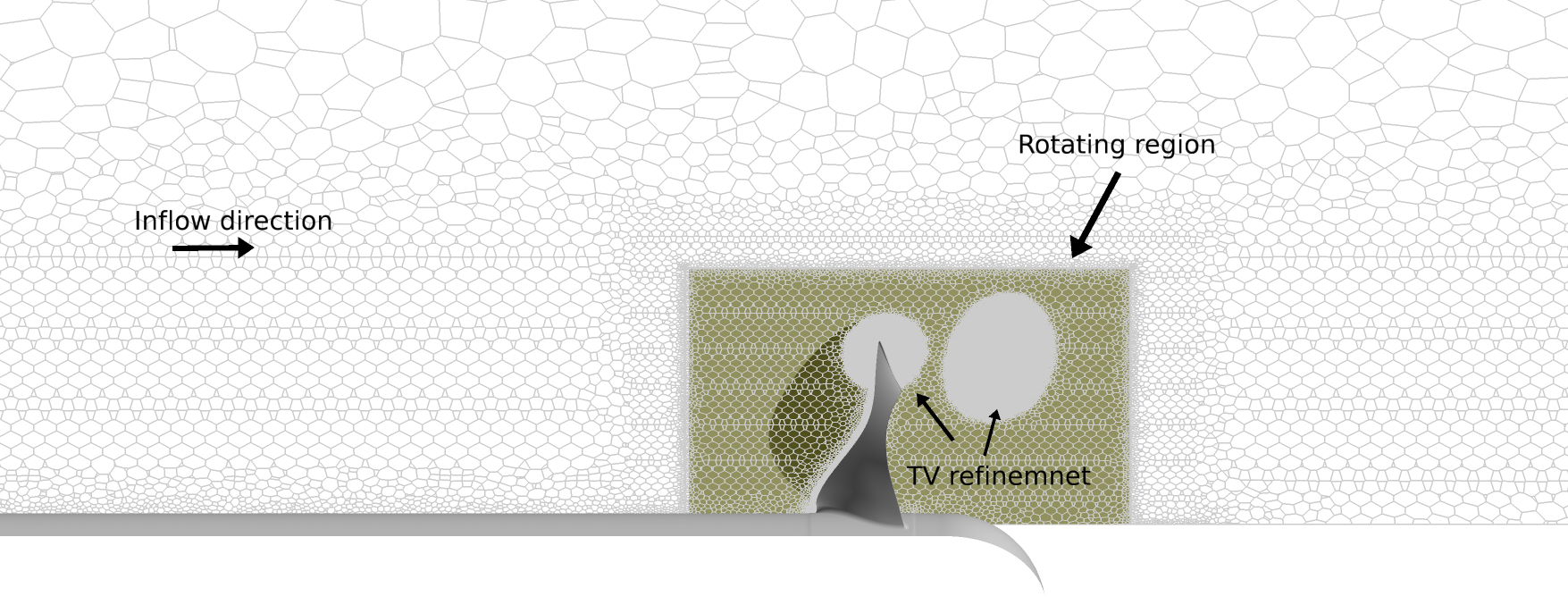}
			\caption{Streamwise resolution}		
			\label{fig::starCCMmeshDisRRProepllerFSMesh}												
        \end{subfigure}
                		               
		\caption{Mesh distribution of the full scale propeller.}
		\label{fig::starCCMmeshDisRRProepllerFS}
\end{figure}

\noindent
The full scale propeller is constructed by the geometrical scale ratio of 15 from the model scale propeller leading to the propeller diameter of 3.8145 m. As noted earlier, the computational domain of the full scale propeller consists of one blade and cyclic boundaries on the sides, Figure \ref{fig::starCCMmeshDisRRProepllerFS}. Similar to the model scale settings, the cells are clustered towards the rotating region. The tip vortex refinements follows the same strategy where three helical refinement regions defined based on the primary vortex trajectory are considered around the blade tip to specify the desired resolution at this region. 

\noindent
In Table \ref{fig::tableFSMSDetails}, the operating conditions and normalised resolution details of the blade and tip vortex refinements are presented. As the inlet velocity is kept constant, the rotational rate is adjusted according to the propeller advance ratio. Similar to the propeller surface resolution, the tip vortex refinement resolutions are presented by the non-dimensionalized terms, i.e. $H_1^{+}=u_\tau \Delta H_1 /\nu$, $H_2^{+}=u_\tau \Delta H_2 /\nu$ and $H_1^{+}=u_\tau \Delta H_3 /\nu$. In these equations, $\Delta H_1$, $\Delta H_2$ and $\Delta H_3$ are the specified cell resolutions in the helical tip refinement regions of H1, H2 and H3.

\begin{table}[h!]
        \centering
        \caption{Operating conditions and mesh specifications of the propeller in model scale (MS) and full scale (FS) conditions.}
\begin{tabular}{|c|c|c|c|c|c|c|c|c|c|c|}
\hline
            & \multicolumn{3}{c|}{\textbf{Propeller setup}} & \multicolumn{3}{c|}{\textbf{Blade Resolution}} & \multicolumn{3}{c|}{\textbf{TV resolution}}             & \multicolumn{1}{c|}{\multirow{2}{*}{\textbf{\begin{tabular}[c]{@{}c@{}}Total number\\ of cells (M)\end{tabular}}}}\\  \cline{2-10}
            & D(m)   &  ${\text{U}_{\text{in}}}$ (m/s)  & Re  &${y^+}$  &\begin{tabular}[c]{@{}c@{}}${x^+}$, ${z^+}$\\ baseline\end{tabular} & \begin{tabular}[c]{@{}c@{}}${x^+}$, ${z^+}$\\ refined tip\end{tabular}     & ${\text{H1}^+}$      & ${\text{H2}^+}$    & ${\text{H3}^+}$  & \multicolumn{1}{l|}{}       \\ \hline
\textbf{MS} & 0.2543        & 4.2            & 1.07$\times10^{6}$       & \multicolumn{1}{c|}{\begin{tabular}[c]{@{}c@{}}5\\ and\\ 35\end{tabular}}                & 250    & 10                 & 40            & 20           & 10  & \multicolumn{1}{c|}{\begin{tabular}[c]{@{}c@{}}87.3\\ and\\ 87.1\end{tabular}}           \\ \hline
\textbf{FS} & 3.8145        & 8              & 30.5$\times10^{6}$       & 50                       & 5000   & 75                & 300           & 150          & 75      & 37.3       \\ \hline
\end{tabular}

		\label{fig::tableFSMSDetails}
\end{table}

\begin{figure}[h!]
        \centering
        \begin{subfigure}[b]{0.25\textwidth}
				\includegraphics[width=\textwidth]{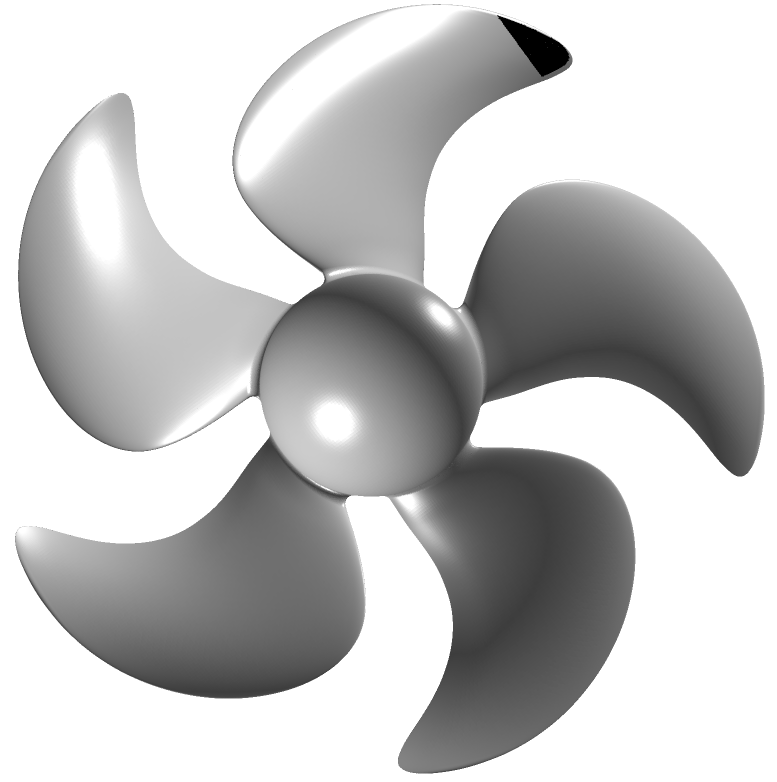}
			\caption{Back side view}
			\label{fig::roughnessAreaBS}		
        \end{subfigure}
		\quad
        \begin{subfigure}[b]{0.25\textwidth}
				\includegraphics[width=0.975\textwidth]{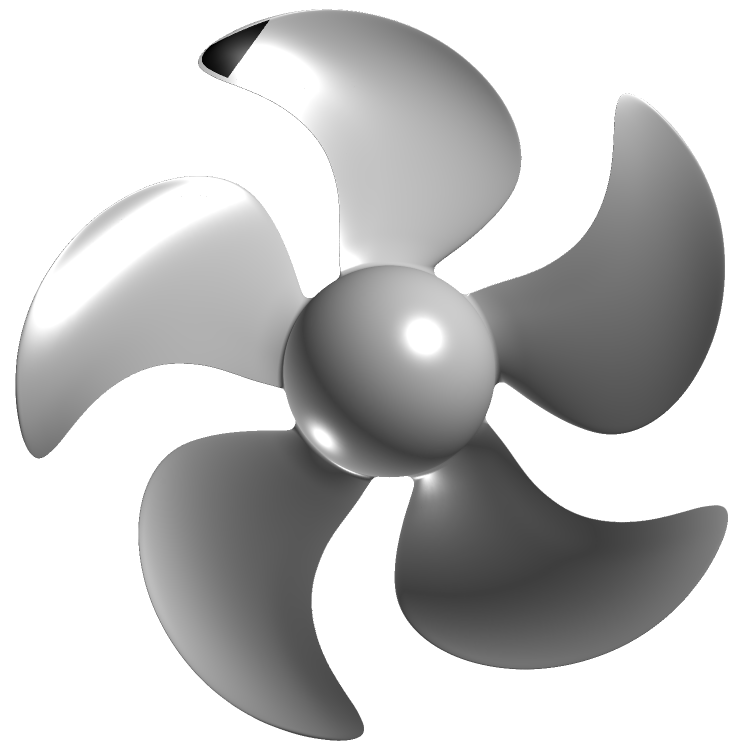}
			\caption{Front side view}
			\label{fig::roughnessAreaFS}
        \end{subfigure}
		\quad
        \begin{subfigure}[b]{0.25\textwidth}
				\includegraphics[width=0.975\textwidth]{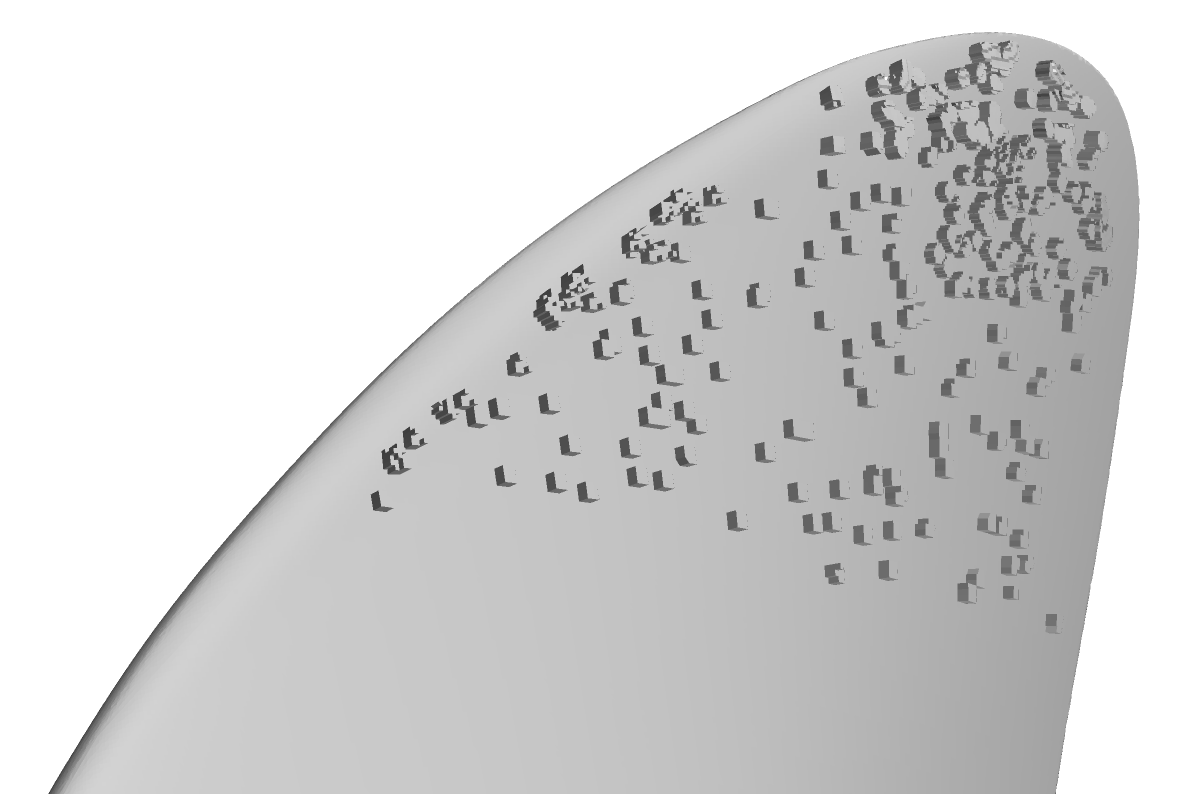}
		\caption{Geometrical roughness}	
		\label{fig::roughnessAreaGeo}											
        \end{subfigure}
                		               
		\caption{(a) and (b): Roughness areas coloured black on the back side and front side of the refined blade; (c): zoomed view of the roughness elements inclusion.}
		\label{fig::roughnessArea}
\end{figure}

\noindent
Based on our previous studies for mitigation of back side tip vortices \cite{asnaghiSMP2019,asnaghiMarine2019}, the tip region of the refined blade, Figure \ref{fig::roughnessArea}, is considered the starting point to investigate the roughness impact on for wider operating conditions. The roughness areas highlighted in Figure \ref{fig::roughnessArea} are thus selected based on the flow properties of tip vortices that incept on the tip or slightly downstream. For the propeller design considered in this study, this type of vortices is observed at low advance ratio numbers, e.g. J=0.82.

\noindent
The study consist of the roughness modelling on the blade tip sides, i.e. the back side and the Front side. For one case where the roughness is only applied on the back side tip region, BS Tip, the mesh topology is modified by removing cells to include the roughness elements into the simulations, Figure \ref{fig::roughnessAreaGeo}. This will provide the opportunity to resolve the flow around these roughness elements.
 
\begin{figure}[h!]
        \centering   
        \begin{subfigure}[b]{0.4\textwidth}	
	        \centering   
				\includegraphics[width=0.8\textwidth]{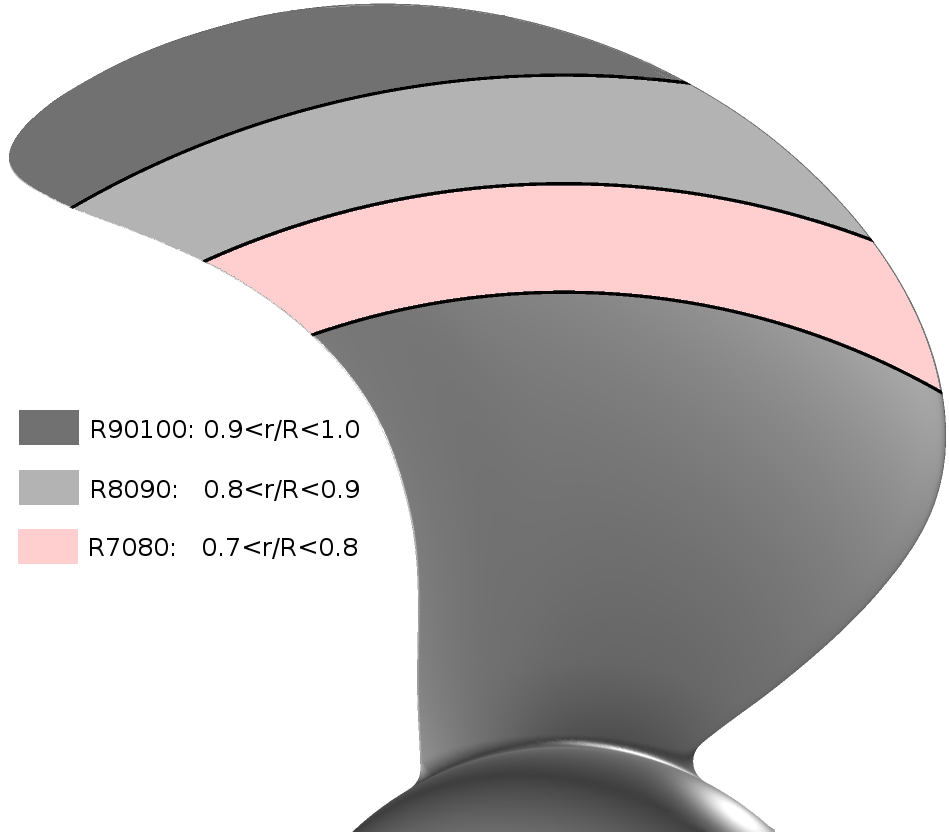}
                \caption{Radial roughness patterns}
                \label{fig::radialFrontSideDistribution}
        \end{subfigure}%
        \quad
        \begin{subfigure}[b]{0.4\textwidth}	
	        \centering   
				\includegraphics[width=0.8\textwidth]{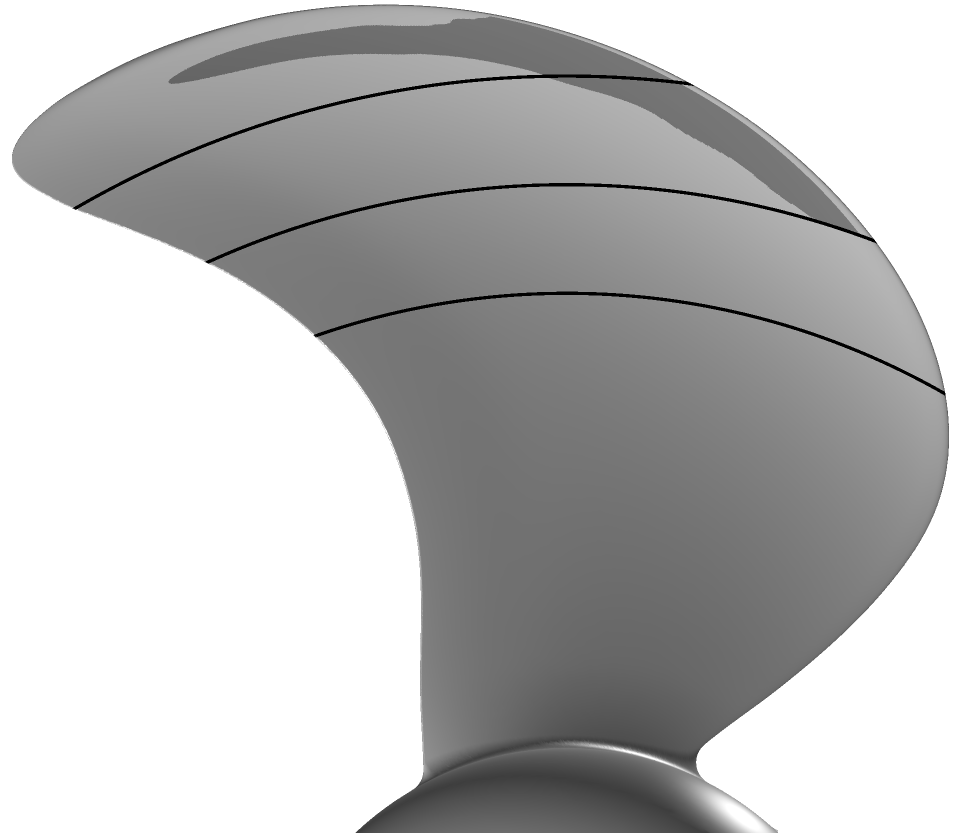}
                \caption{Radial leading edge roughness pattern}
                \label{fig::RE80100Distribution}
        \end{subfigure}%
                                  
		\caption{Roughness patterns tested on the front side of the propeller.}
		\label{fig::FrontSideroughnessSpecifications}
\end{figure}

\noindent
At higher J values, e.g. J=1.26, the main vortex appears as a leading edge vortex formed on the front side of the blade, and it is thus expected the area covered with roughness is different. Therefore, the roughness pattern optimization for this vortex consists of investigation of radial areas defined in Figure \ref{fig::radialFrontSideDistribution}. In order to find which part of radial areas will have more impact on TV mitigation, also smaller areas along the leading edge are considered, e.g. as in Figure \ref{fig::RE80100Distribution}.

\noindent
The summary of evaluated roughness patterns and arrangements are presented in Table \ref{fig::tableroughnessPatternSummary}.
\begin{table}[h!]
        \centering
        \caption{Summary of the roughness patterns tested on the back and front sides of the blade.}
\begin{tabular}{|l|l|c|}
\hline
\textbf{Pattern} & \textbf{Where the roughness is applied:}                                                & \multicolumn{1}{l|}{\textbf{Related figure}} \\ \hline
Smooth                     & The blade is smooth.                                                                    & \multirow{4}{*}{-}                           \\ \cline{1-2}
FS                         & Front side of the blade                                                                 &                                              \\ \cline{1-2}
BS                         & Back side of the blade                                                                  &                                              \\ \cline{1-2}
FR                         & Both sides of the blade (fully roughened)                                               &                                              \\ \hline \hline
BS Tip                     & Back side tip                                                                           & Figure \ref{fig::roughnessAreaBS}                                       \\ \hline
FS Tip                     & Front side tip                                                                          & Figure \ref{fig::roughnessAreaFS}                                       \\ \hline
BS + FS Tip                & Tip of the back and front sides                                                         & Figure \ref{fig::roughnessAreaBS} + \ref{fig::roughnessAreaFS}                                       \\ \hline \hline
R7080                      & Radial distance 0.7\textless{}r/R\textless{}0.8 on the front side                       & \multirow{4}{*}{Figure \ref{fig::radialFrontSideDistribution}}                      \\ \cline{1-2}
R8090                      & Radial distance 0.8\textless{}r/R\textless{}0.9 on the front side                       &                                              \\ \cline{1-2}
R90100                     & Radial distance 0.9\textless{}r/R\textless{}1.0 on the front side                       &                                              \\ \cline{1-2}
R70100                     & Radial distance 0.7\textless{}r/R\textless{}1.0 on the front side                       &                                              \\ \hline \hline
RE8090                     & Leading edge on the radial distance 0.8\textless{}r/R\textless{}0.9 of the front side   & \multirow{4}{*}{Figure \ref{fig::RE80100Distribution}}                      \\ \cline{1-2}
RE8595                     & Leading edge on the radial distance 0.85\textless{}r/R\textless{}0.95 of the front side &                                              \\ \cline{1-2}
RE90100                    & Leading edge on the radial distance 0.9\textless{}r/R\textless{}1.0 of the front side   &                                              \\ \cline{1-2}
RE80100                    & Leading edge on the radial distance 0.8\textless{}r/R\textless{}1.0 of the front side   &                                              \\ \hline \hline
ORP                    & Optimum roughness pattern = combination of BS Tip and RE80100 &    Figure \ref{fig::roughnessAreaBS} + \ref{fig::RE80100Distribution}                                          \\ \hline
\end{tabular}
		\label{fig::tableroughnessPatternSummary}
\end{table}
Following our previous studies \cite{asnaghiSMP2019,asnaghiMarine2019}, all of the analysis is performed by considering a fixed value for the roughness height, K$_s$=250 $\mu$m in model scale condition; this corresponds to $\text{K}_\text{s}^{+}=35$. The roughness height is extended into the full scale condition by considering the geometrical scale ratio. This gives K$_s$=3.75 mm in full scale conditions corresponding to $\text{K}_\text{s}^{+} \approx 925$.     
\section{Results}
\begin{figure}[h!]
    \centering
					\begin{tikzpicture}
					\pgfplotsset{xmin=0.8, xmax=1.3,width=0.8\textwidth, legend pos= north east, legend columns=3}
						\begin{axis}[axis y line*=right,ymin=0, ymax=1.0,  xlabel=J,  ylabel=$\eta_0$]
						
							\addplot[only marks,mark=diamond*] table[only marks, x index=0,y index=3] 	{PLOT/1301B-OW/kT_Exp.dat};  
							\label{Exp-eff-1301B}
							\addlegendentry{Exp-$\eta_0$-MS}

							\addplot[smooth,orange] table[x index=0,y index=3] 	{PLOT/1301B-OW/MS_kT_Num.dat}; 
							\label{Num-eff-1301B-MS}
							\addlegendentry{Num-$\eta_0$-MS}
							
							\addplot[dashed,orange] table[x index=0,y index=3] 	{PLOT/1301B-OW/FS_kT_Num.dat}; 
							\label{Num-eff-1301B-FS}
							\addlegendentry{Num-$\eta_0$-FS}
							
						\end{axis}
					
						\begin{axis}[axis y line*=left,  axis x line=none,  xlabel=J,  ymin=0, ymax=0.8,  ylabel=$k_T$ and $10k_Q$]
							\addlegendimage{/pgfplots/refstyle=Exp-eff-1301B}\addlegendentry{Exp-$\eta_0$-MS}

							\pgfplotsset{every x tick label/.append style={font=\large, yshift=-3ex}}
							\pgfplotsset{every y tick label/.append style={font=\large, xshift=-1ex}}
			
							\addplot[only marks,mark=square*,black] table[only marks, x index=0,y index=1] 	{PLOT/1301B-OW/kT_Exp.dat};  
							\addlegendentry{Exp-$K_T$-MS}

							\addplot[only marks,mark=*,black] table[only marks, x index=0,y index=2] 	{PLOT/1301B-OW/kT_Exp.dat};
						 	\addlegendentry{Exp-$K_Q$-MS}

							\addlegendimage{/pgfplots/refstyle=Num-eff-1301B-MS}\addlegendentry{Num-$\eta_0$-MS}
							
							\addplot[smooth,blue] table[x index=0,y index=1] 	{PLOT/1301B-OW/MS_kT_Num.dat};
							\addlegendentry{Num-$K_T$-MS}

							\addplot[smooth,red] table[x index=0,y index=2] 	{PLOT/1301B-OW/MS_kT_Num.dat};
							\addlegendentry{Num-$K_Q$-MS}	
							
							\addlegendimage{/pgfplots/refstyle=Num-eff-1301B-FS}\addlegendentry{Num-$\eta_0$-FS}
										
							\addplot[dashed,blue] table[x index=0,y index=1] 	{PLOT/1301B-OW/FS_kT_Num.dat};
							\addlegendentry{Num-$K_T$-FS}

							\addplot[dashed,red] table[x index=0,y index=2] 	{PLOT/1301B-OW/FS_kT_Num.dat};
							\addlegendentry{Num-$K_Q$-FS}	
							
						\end{axis}
						
					\end{tikzpicture}

    \caption{Open water performance of the smooth propeller in model scale and full scale conditions.}
    \label{fig:MSFSOWForcesCoeff}
\end{figure}
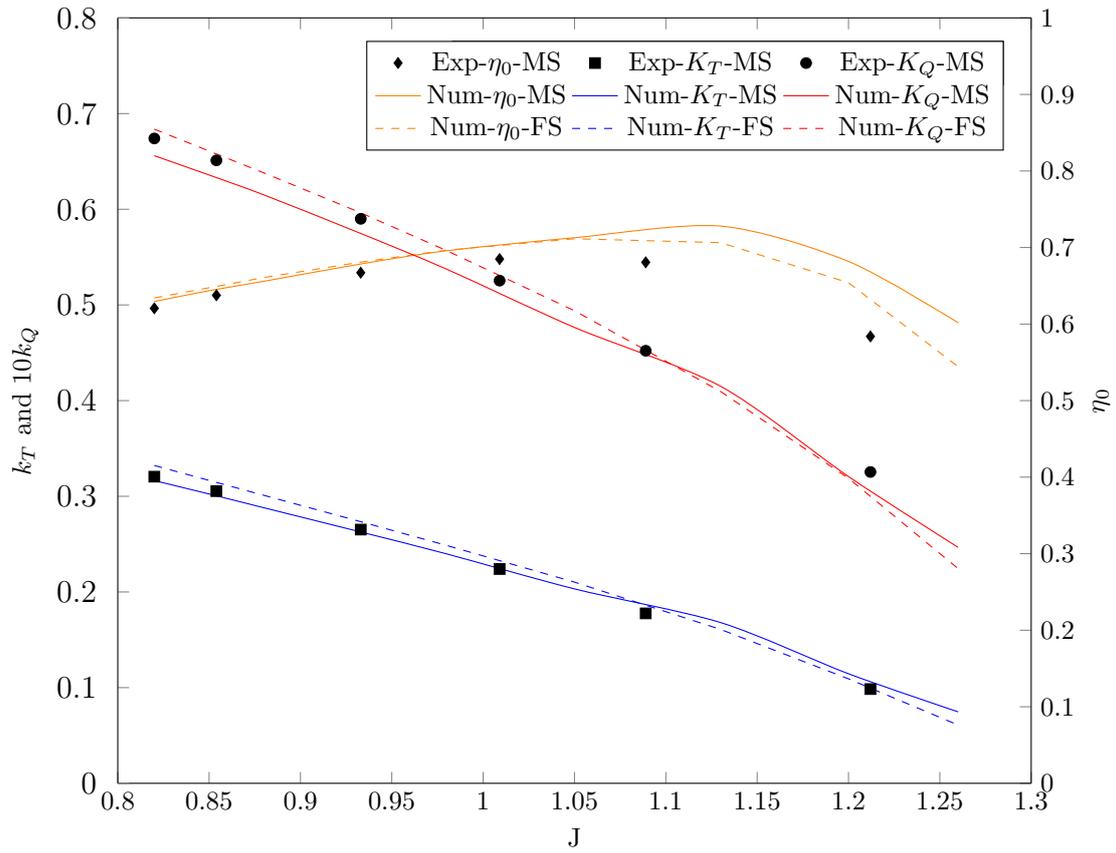

\noindent
Open water performance of the propeller in the model scale and full scale conditions are presented in Figure \ref{fig:MSFSOWForcesCoeff}. The figure includes thrust coefficient, $K_T$, torque coefficient, $K_Q$, and efficiency, $\eta_0$, at different advance ratio values, J. The model scale results are presented by solid lines while the full scale results are shown by dashed lines. 

\noindent
General accuracy of the results is satisfactory and agrees well with the experimental measurements. The trend in the forces, however, are predicted differently in model and full scales. In J$<$1.1, the thrust and torque coefficients of the full scale condition are predicted higher than the model scale predictions. This is found to make a small difference on the predicted efficiency in this range as the increase of both coefficients cancels each other. In higher J values, i.e. 1.1$<$J, the trend of thrust and torque predictions between model scale and full scale switches where both of forces are underpredicted in the full scale condition, although this is more apparent for the thrust. This leads to predictions of lower efficiency in full scale compared to the model scale results.

\begin{figure}[h!]
    \centering

	\begin{tikzpicture}
			\begin{axis}[font=\large,legend pos=north east, legend columns=4,  xlabel=J, ylabel=$\sigma_i$, width=0.74\textwidth, xmin=0.8, xmax=1.3, ymin=0, ymax=10, xtick={0.8,0.9,1.0,1.1,1.2,1.3}, ytick={0,1,2,3,4,5,6,7,8}]
			  
			\pgfplotsset{every x tick label/.append style={font=\large, yshift=-0.75ex}}
			\pgfplotsset{every y tick label/.append style={font=\large, xshift=-0.75ex}}
			  
			\addplot[only marks,mark=*,blue] 	table[only marks, x index=0,y index=1] 		{PLOT/1301B-OW/InceptionExp/BBC_Exp.dat};  
			\label{Exp-BBC}
			\addlegendentry{\footnotesize  Exp-BBC}

			\addplot[smooth,blue] 	table[x index=0,y index=1] 		{PLOT/1301B-OW/InceptionExp/BBC_ExpExtrapolated.dat};  
			\label{Exp-BBC}
			\addlegendentry{\footnotesize   ExpExtrapolated-BBC}

			\addplot[only marks,mark=triangle,blue, mark size=3pt] 	table[only marks, x index=0,y index=1] 		{PLOT/1301B-OW/InceptionMinPressure/BBC_minPressure.dat};  
			\label{BBC-MS}
			\addlegendentry{\footnotesize  BBC-MS}
			
			\addplot[only marks,mark=triangle*,blue, mark size=3pt] 	table[only marks, x index=0,y index=1] 		{PLOT/1301B-OW/InceptionMinPressureFS/BBC_minPressure.dat};  
			\label{BBC-FS}
			\addlegendentry{\footnotesize  BBC-FS}		
			
			\addplot[only marks,mark=*,orange] 	table[only marks,x index=0,y index=1] 		{PLOT/1301B-OW/InceptionExp/BTV_Exp.dat}; 
			\label{Exp-BTV}
			\addlegendentry{\footnotesize  Exp-BTV}
			
			\addplot[smooth,orange] 	table[x index=0,y index=1] 		{PLOT/1301B-OW/InceptionExp/BTV_ExpExtrapolated.dat}; 
			\label{Exp-BTV}
			\addlegendentry{\footnotesize  ExpExtrapolated-BTV}
		
			\addplot[only marks,mark=triangle,orange, mark size=3pt] 	table[only marks,x index=0,y index=1] 		{PLOT/1301B-OW/InceptionMinPressure/BTV_minPressure.dat}; 
			\label{BTV-MS}
			\addlegendentry{\footnotesize  BTV-MS}		

			\addplot[only marks,mark=triangle*,orange, mark size=3pt] 	table[only marks,x index=0,y index=1] 		{PLOT/1301B-OW/InceptionMinPressureFS/BTV_minPressure.dat}; 
			\label{BTV-FS}
			\addlegendentry{\footnotesize  BTV-FS}								
			
			\addplot[only marks,mark=*,green] 	table[only marks, x index=0,y index=1] 		{PLOT/1301B-OW/InceptionExp/FTV_Exp.dat};
			\label{Exp-FTC}
			\addlegendentry{\footnotesize  Exp-FTV}
	
			\addplot[smooth,green] 	table[x index=0,y index=1] 		{PLOT/1301B-OW/InceptionExp/FTV_ExpExtrapolated.dat};
			\label{Exp-FTC}
			\addlegendentry{\footnotesize  ExpExtrapolated-FTV}	

			\addplot[only marks,mark=triangle,green, mark size=3pt] 	table[only marks, x index=0,y index=1] 		{PLOT/1301B-OW/InceptionMinPressure/FTV_minPressure.dat};
			\label{FTV-MS}
			\addlegendentry{\footnotesize  FTV-MS}	

			\addplot[only marks,mark=triangle*,green, mark size=3pt] 	table[only marks, x index=0,y index=1] 		{PLOT/1301B-OW/InceptionMinPressureFS/FTV_minPressure.dat};
			\label{FTV}
			\addlegendentry{\footnotesize  FTV-FS}	
																						
		\end{axis}

	\end{tikzpicture}

    \caption{Open water cavitation inception diagram of the smooth propeller in the model scale (MS) and full scale (FS) conditions represented by FTV: front tip vortex, BTV: back tip vortex, and BBC: back bubble cavitation.}
    \label{fig:inceptionDiagram1301BOWMSFS}
\end{figure}
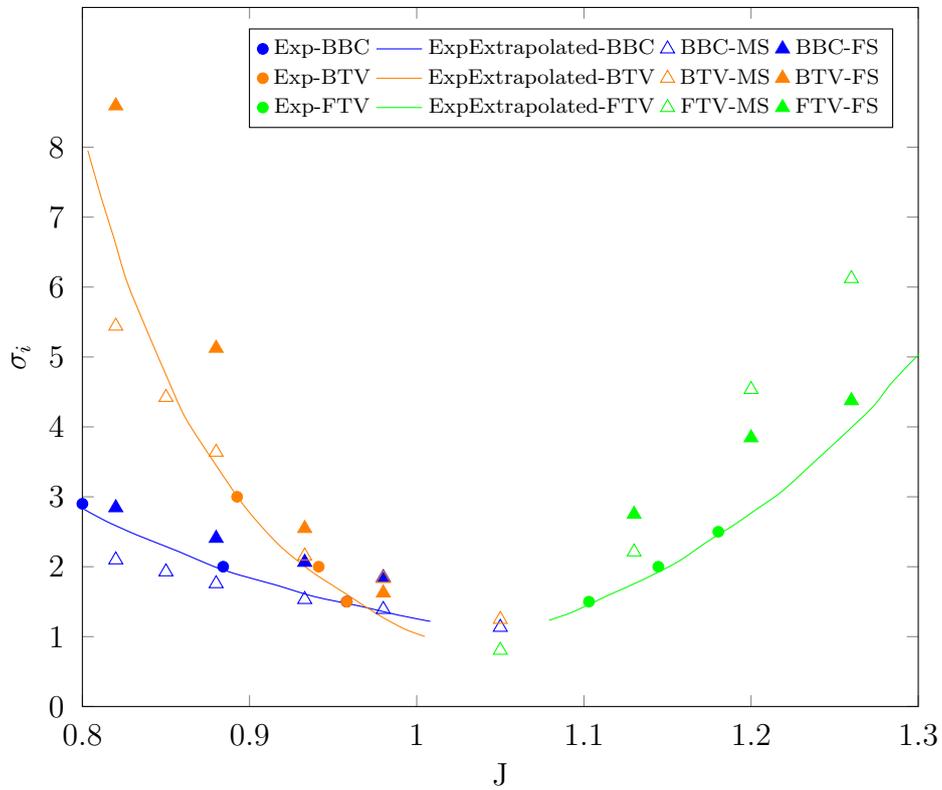

\noindent
The cavitation inception prediction of the propeller for the model scale and full scale conditions are presented in Figure \ref{fig:inceptionDiagram1301BOWMSFS}. Three different cavitation types are investigated: FTV, front tip vortex; BTV, back tip vortex; and BBC, back bubble cavitation. In this figure, the experimental measurements are presented by solid circular symbols while numerical results predicted by the minimum pressure criterion is presented by triangular symbols. In order to provide further data for analysis, the experimental measurements are extrapolated to the sides of the chart, presented by solid lines. 

\noindent
The overall comparison in model scale is quite satisfactory, with BTV well predicted and BBC only somewhat smaller inception values, while the FTV shows larger discrepancies; however the minimum point in the bucket diagram is well captured. Comparison of the model scale and full scale results indicate a stronger back side tip vortex and bubble cavitation predictions in the full scale condition. The prediction of the front side tip vortex is found to be opposite where a stronger front tip vortex is observed in the model scale condition. 

\subsection{Flow properties of model scale propeller}
\begin{figure}[h!]
        \centering 
        \quad                
        \begin{subfigure}[b]{0.95\textwidth}			
			\begin{overpic}[width=0.3\textwidth]{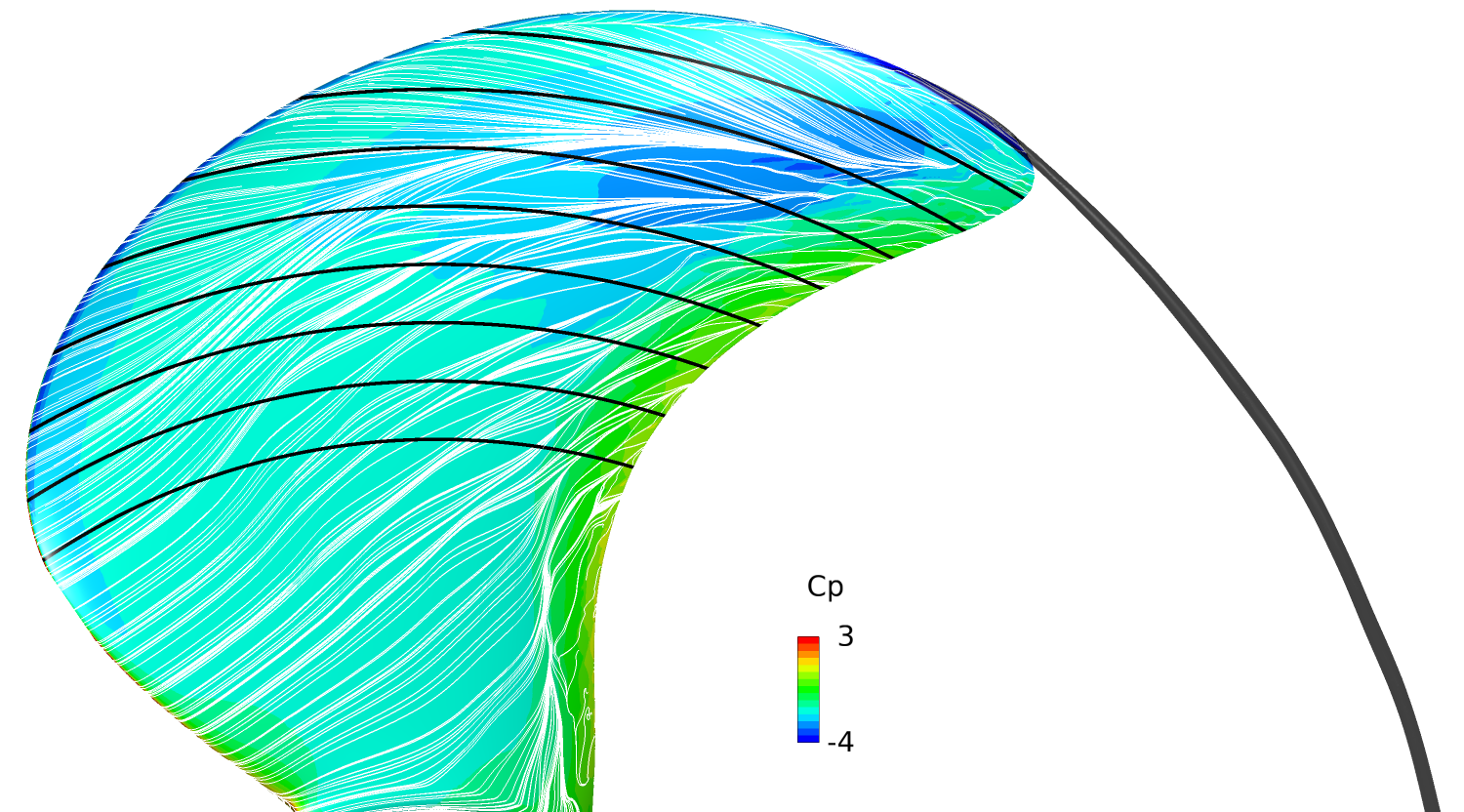} 
				\put (60,60) {\makebox(0,0)[br]{\textcolor{black}{J=0.820}}}	
				\put (83,16) {\makebox(0,0)[br]{\textcolor{black}{r/R=0.7}}}	
				\put(60,24.5){\vector(-2,1){10}}							
			\end{overpic}
			\begin{overpic}[width=0.31\textwidth]{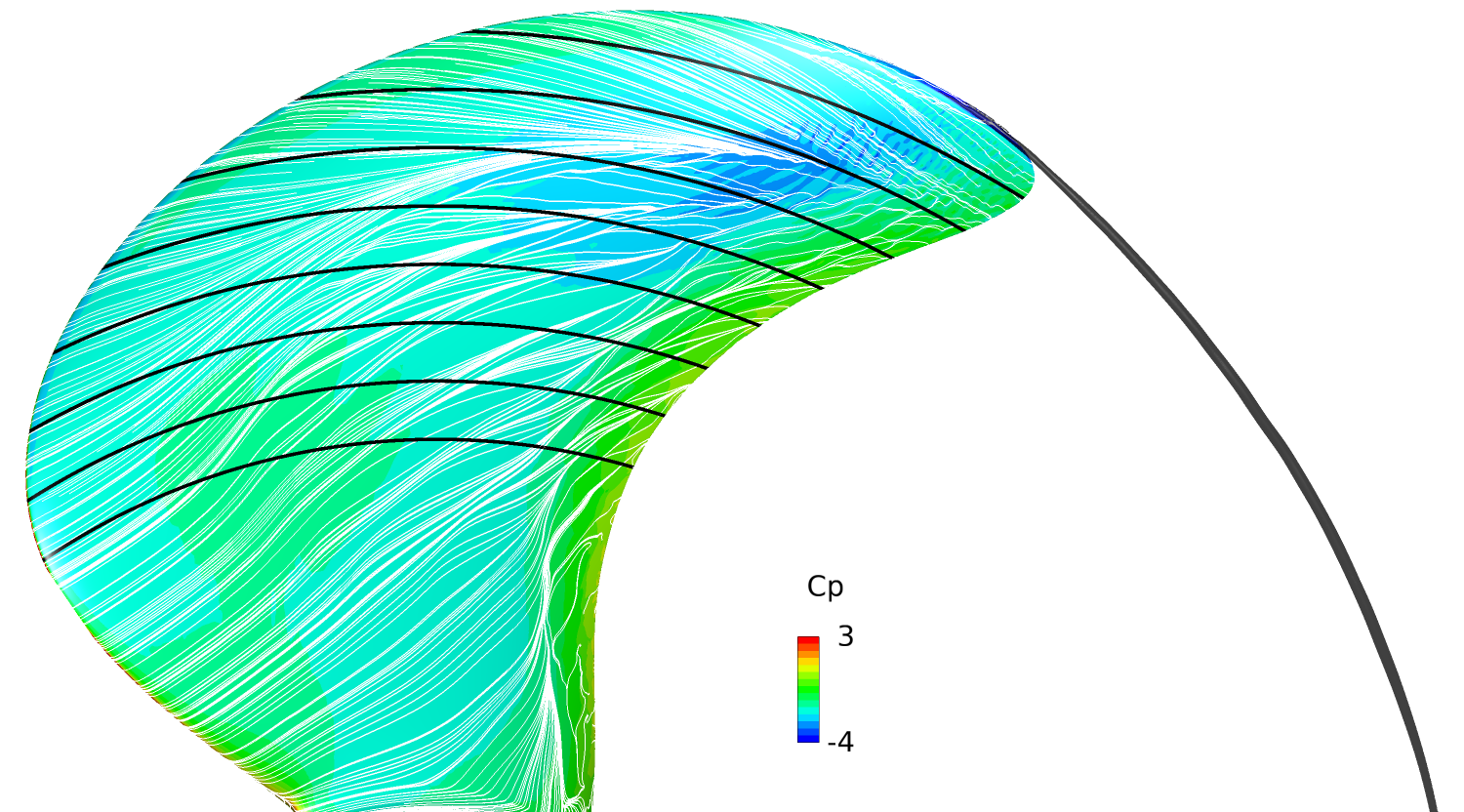} 
				\put (60,60) {\makebox(0,0)[br]{\textcolor{black}{J=0.854}}}			
			\end{overpic}
			\begin{overpic}[width=0.31\textwidth]{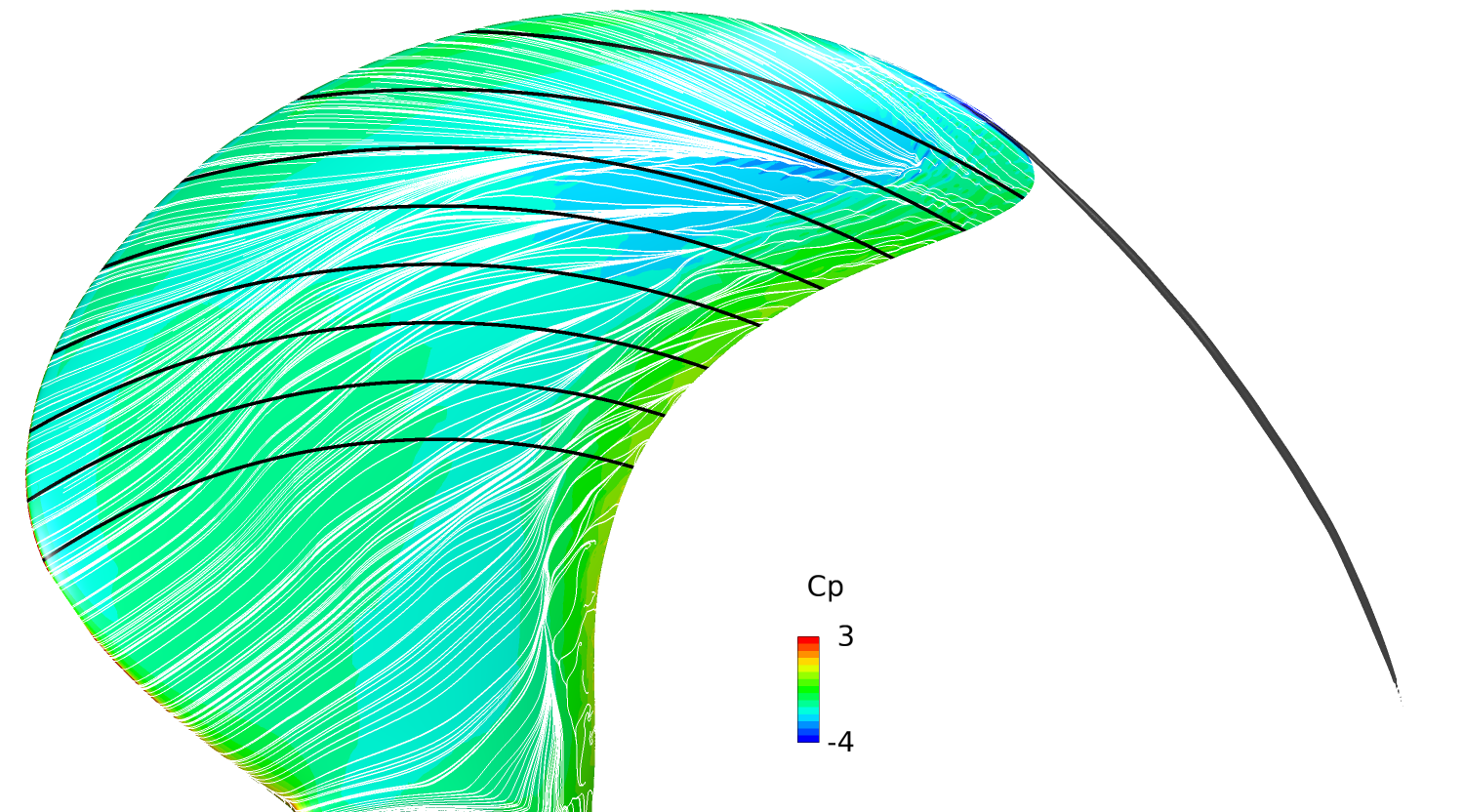}
				\put (60,60) {\makebox(0,0)[br]{\textcolor{black}{J=0.873}}}			
			\end{overpic}	
			\label{fig:smoothPropellerFPStreamlines1backSide}
            \caption{Back side}						   
        \end{subfigure}
        \quad        
        \begin{subfigure}[b]{0.95\textwidth}			
			\begin{overpic}[width=0.3\textwidth]{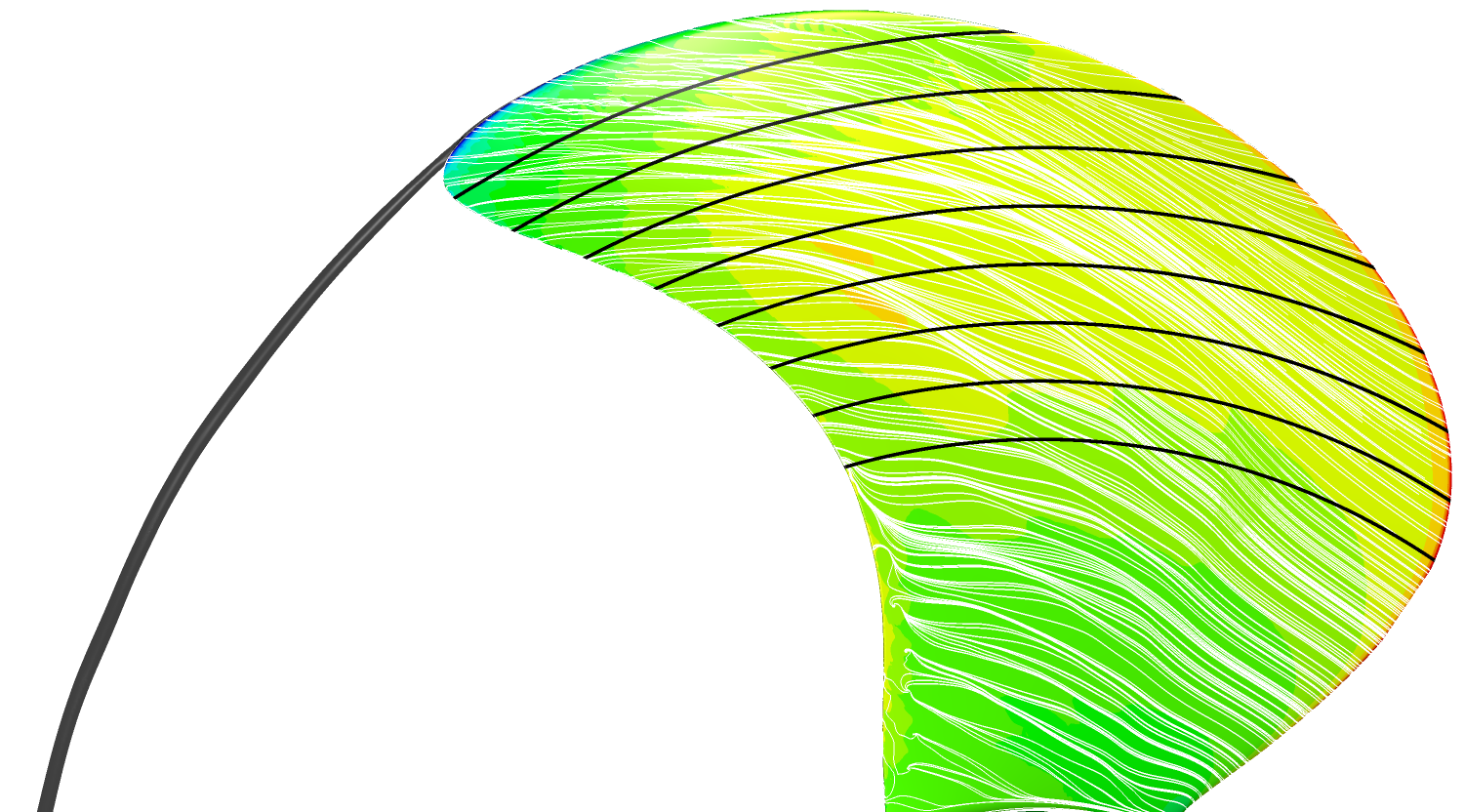} 
				\put (50,15) {\makebox(0,0)[br]{\textcolor{black}{r/R=0.7}}}	
				\put(42.5,23.2){\vector(3,2){10}}	
			\end{overpic}
			\begin{overpic}[width=0.31\textwidth]{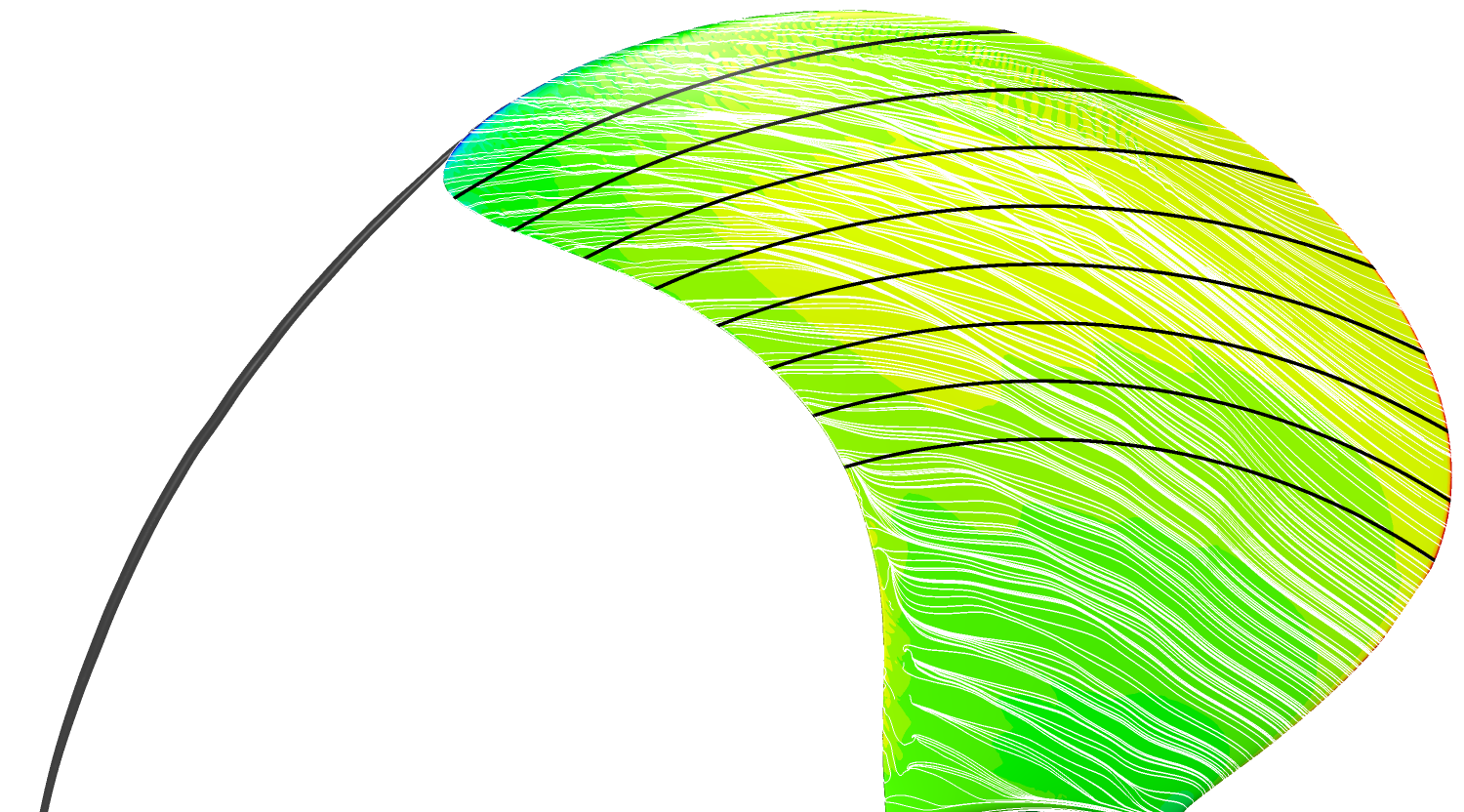} 
			\end{overpic}
			\begin{overpic}[width=0.31\textwidth]{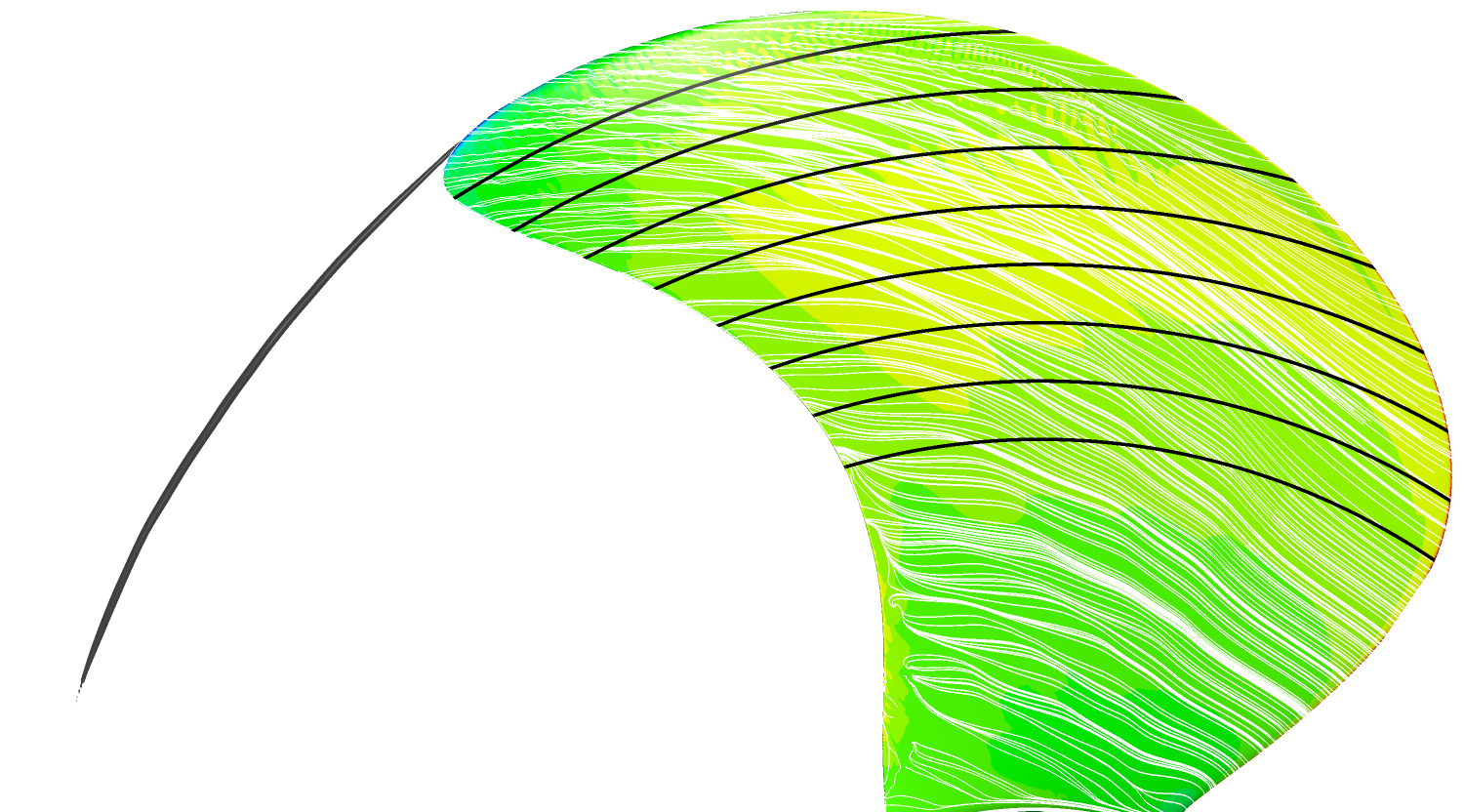}
			\end{overpic}
			\label{fig::smoothPropellerFPStreamlines1PS}			
			 \caption{Front side}							    
        \end{subfigure}                          
            \caption{Flow properties at different back side tip vortex conditions. The blade surface is coloured by $\text{C}_\text{p}$ distribution, the flow streamlines are presented in white, the tip vortex is presented by the pressure iso-surface equal to $\text{C}_\text{p}=-4$ coloured in black.}
		\label{fig::smoothPropellerFPStreamlines1}
\end{figure}

 \noindent 
The flow properties at low J values where the tip vortex forms on the back side of the blade are presented in Figure \ref{fig::smoothPropellerFPStreamlines1}. In this figure, the tip vortex is presented by the pressure coefficient iso-surface, the flow streamlines are presented by white lines, and the blade surface is coloured by the pressure coefficient distribution.

\noindent 
As expected, at lower J values, which corresponds to higher propeller rotational speeds, the back tip vortex becomes stronger. This corresponds to more concentrated flow streamlines which feed the tip vortex momentum. This can be noted on the back side especially close to the trailing edge in 0.9$<$r/R$<$0.95. On the front side, no obvious concentrated streamlines in any region are observed. It seems that the flow streamlines from the front side evenly contribute to the tip vortex roll-up downstream of the tip rather than where the tip vortex starts.
\vspace{0.5 cm}

\begin{figure}[h!]
        \centering   
        \begin{subfigure}[b]{0.95\textwidth}			
			\begin{overpic}[width=0.3\textwidth]{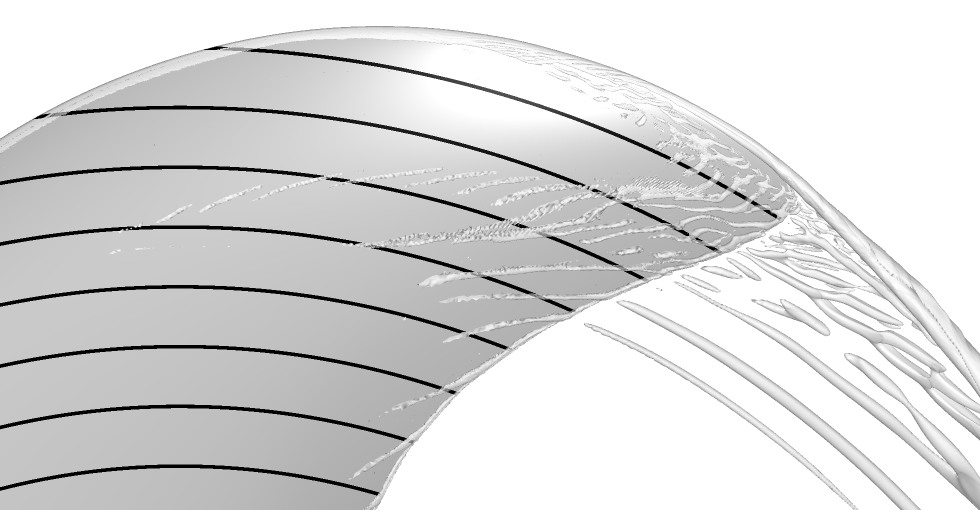} 
				\put (60,50) {\makebox(0,0)[br]{\textcolor{black}{J=0.820}}}
				\put (83,-2) {\makebox(0,0)[br]{\textcolor{black}{r/R=0.7}}}	
				\put(58,7){\vector(-2,1){10}}		
			\end{overpic}
			\begin{overpic}[width=0.31\textwidth]{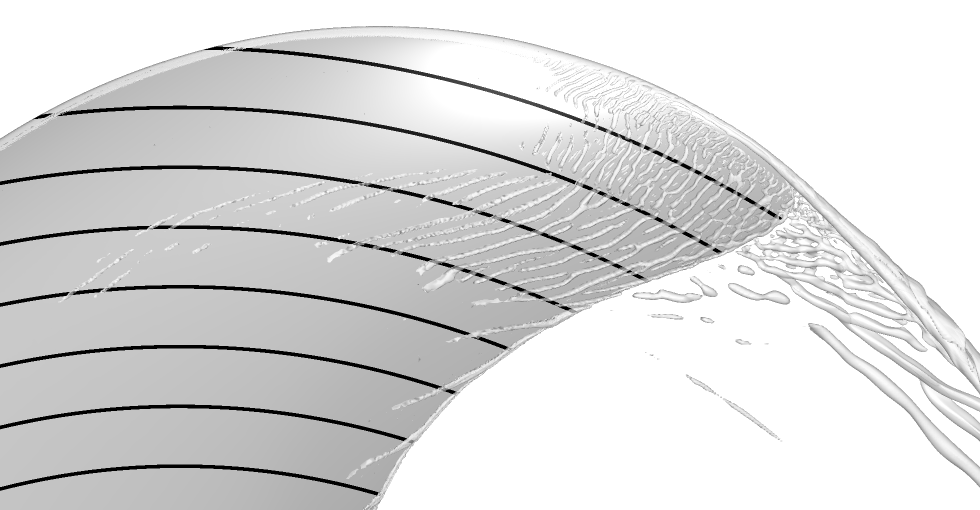} 
				\put (60,50) {\makebox(0,0)[br]{\textcolor{black}{J=0.854}}}
			\end{overpic}
			\begin{overpic}[width=0.31\textwidth]{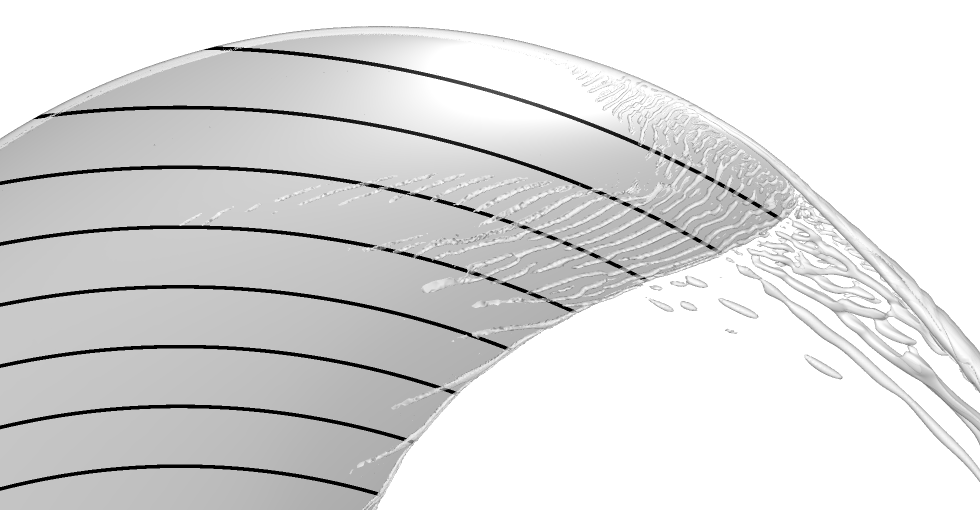} 
				\put (60,50) {\makebox(0,0)[br]{\textcolor{black}{J=0.873}}}
			\end{overpic}
			\label{fig::smoothPropellerFPQ1SS}										    
            \caption{Zoomed-view back side}
        \end{subfigure}   

        \begin{subfigure}[b]{0.95\textwidth}			
			\begin{overpic}[width=0.3\textwidth]{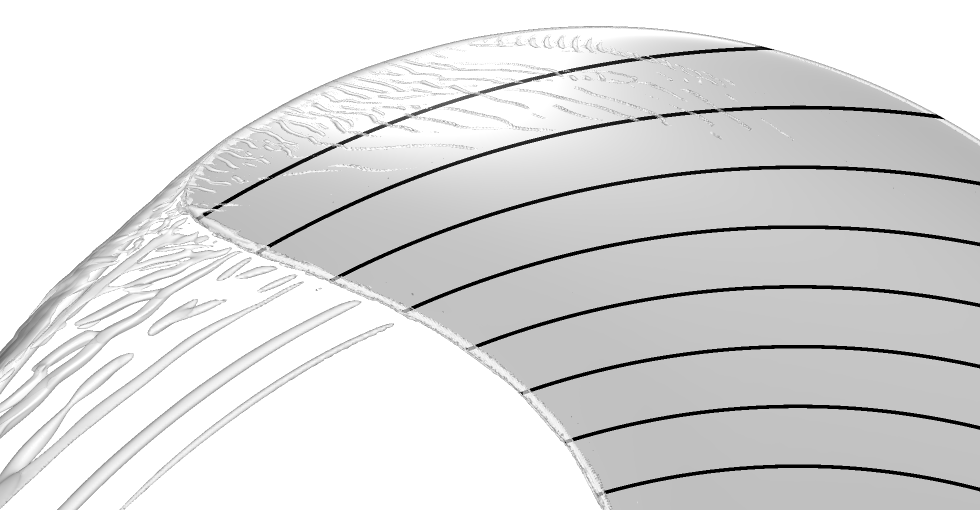} 
				\put (43,-2) {\makebox(0,0)[br]{\textcolor{black}{r/R=0.7}}}	
				\put (41,6.2){\vector(2,1){10}}	
			\end{overpic}
			\begin{overpic}[width=0.31\textwidth]{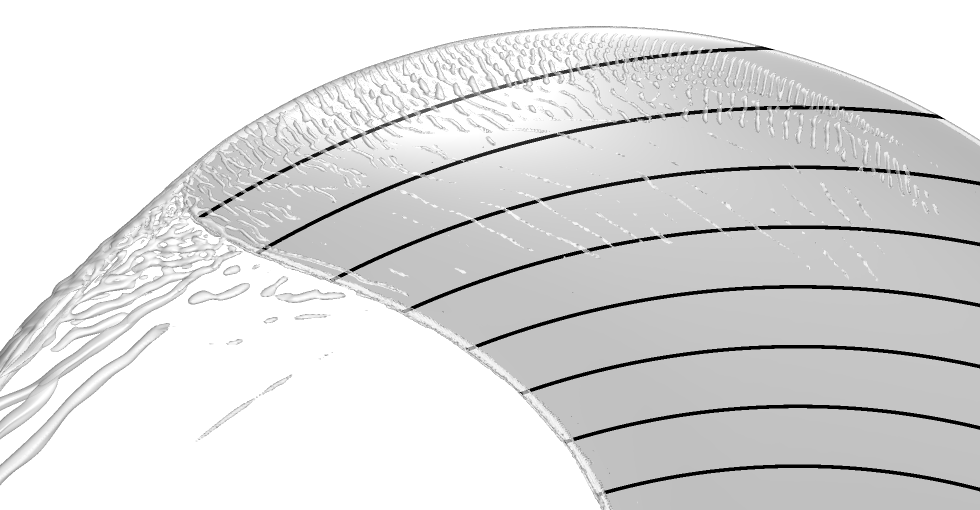} 
			\end{overpic}
			\begin{overpic}[width=0.31\textwidth]{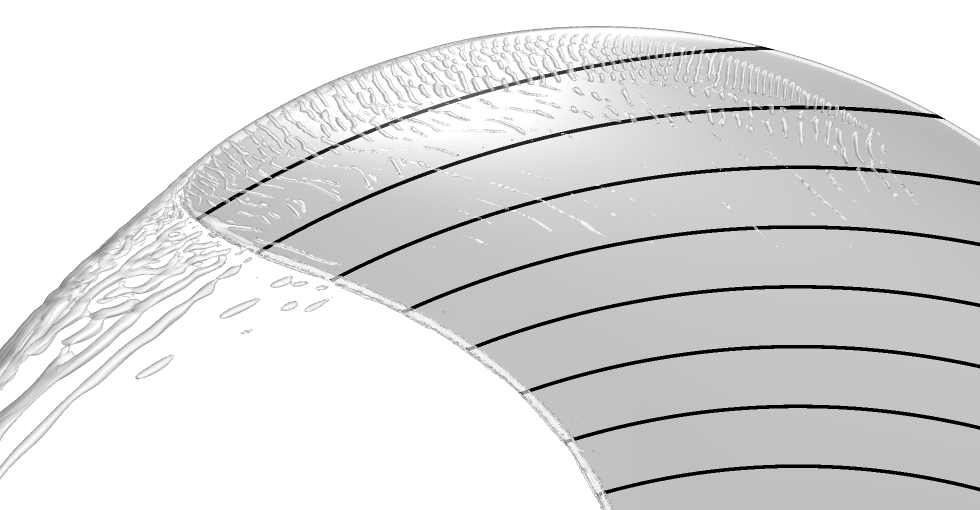} 
			\end{overpic}
			\label{fig::smoothPropellerFPQ1PS}										    
            \caption{Zoomed-view front side}
        \end{subfigure}  
                          
		\caption{Distribution of Q-criterion iso-surface = 200 around the blade tip at low J values.}
		\label{fig::smoothPropellerFPQ1}
\end{figure}

\noindent 
In Figure \ref{fig::smoothPropellerFPQ1}, distributions of vortical structures at low J values are presented by the iso-surface of Q-criterion equal to 200. The results clearly show that at higher rotational rate, i.e. lower J values, more trailing vortices are formed on the blade close to the trailing edge, especially after the blade mid-chord. More interestingly, in 0.9$<$r/R$<$1.0, the vortical structures distinguish a triangular area on the tip which corresponds to the area where concentrated flow streamlines enter. 

\noindent 
Based on the flow properties analysis of lower J values, triangular areas on the back side and front side of the blade are considered as effective areas in formation of the back side tip vortex. These areas are presented in Figure \ref{fig::roughnessArea}.

\begin{figure}[h!]
        \centering 
        \begin{subfigure}[b]{0.95\textwidth}			
			\begin{overpic}[width=0.3\textwidth]{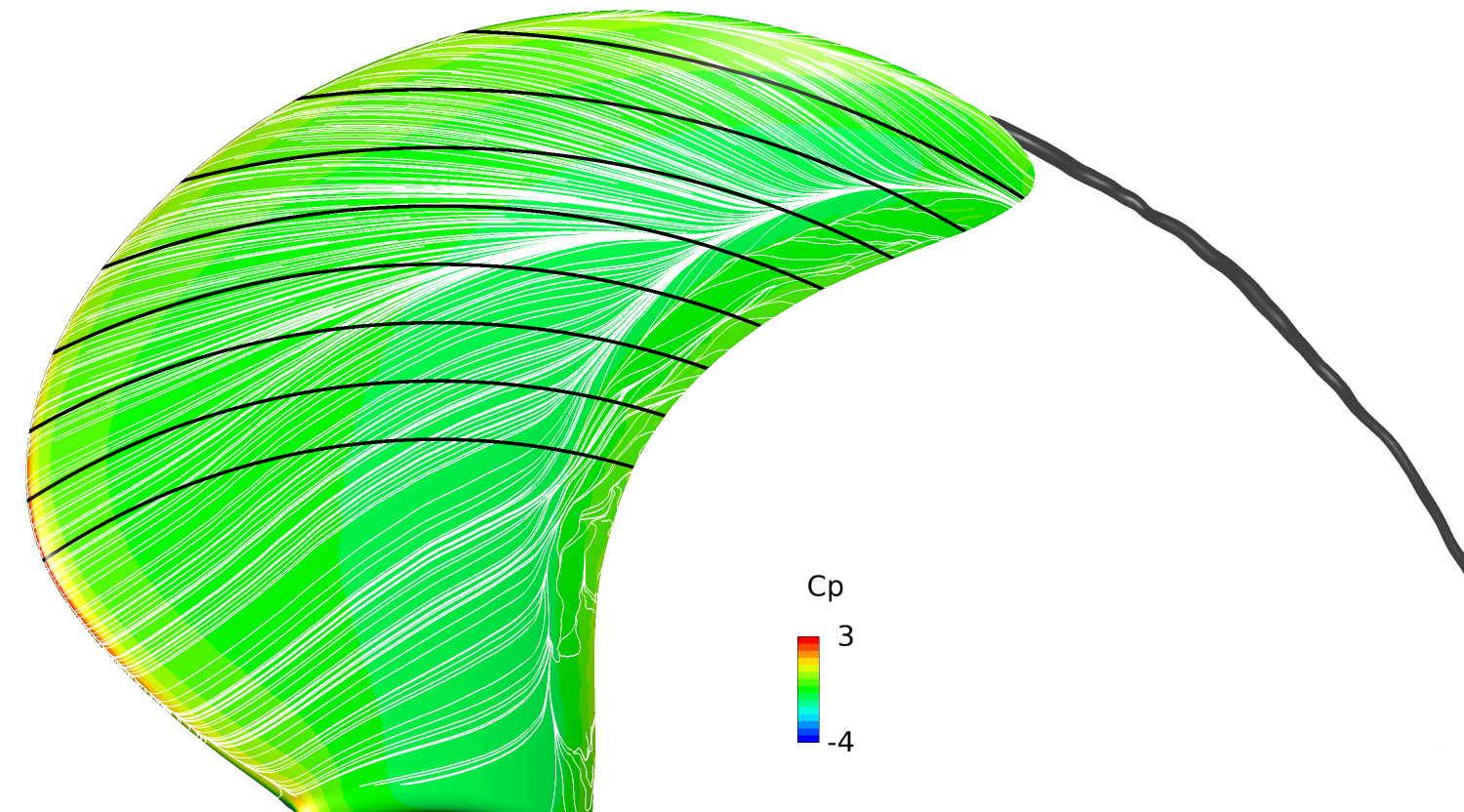} 
				\put (60,60) {\makebox(0,0)[br]{\textcolor{black}{J=1.260}}}	
				\put (88,18.5) {\makebox(0,0)[br]{\textcolor{black}{r/R=0.7}}}	
				\put(60,24){\vector(-2,1){10}}								
			\end{overpic}
			\begin{overpic}[width=0.31\textwidth]{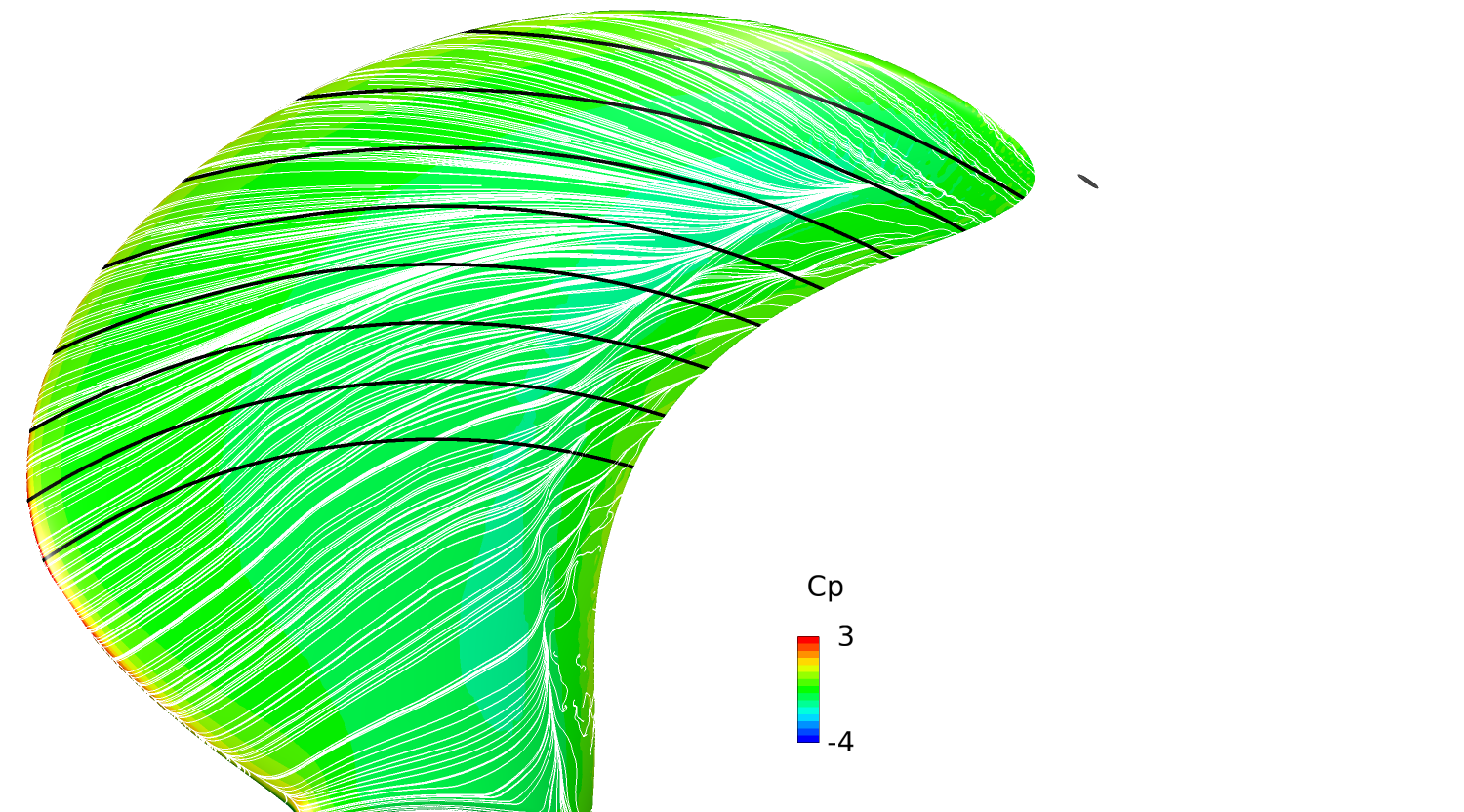} 
				\put (60,60) {\makebox(0,0)[br]{\textcolor{black}{J=1.152}}}			
			\end{overpic}
			\begin{overpic}[width=0.31\textwidth]{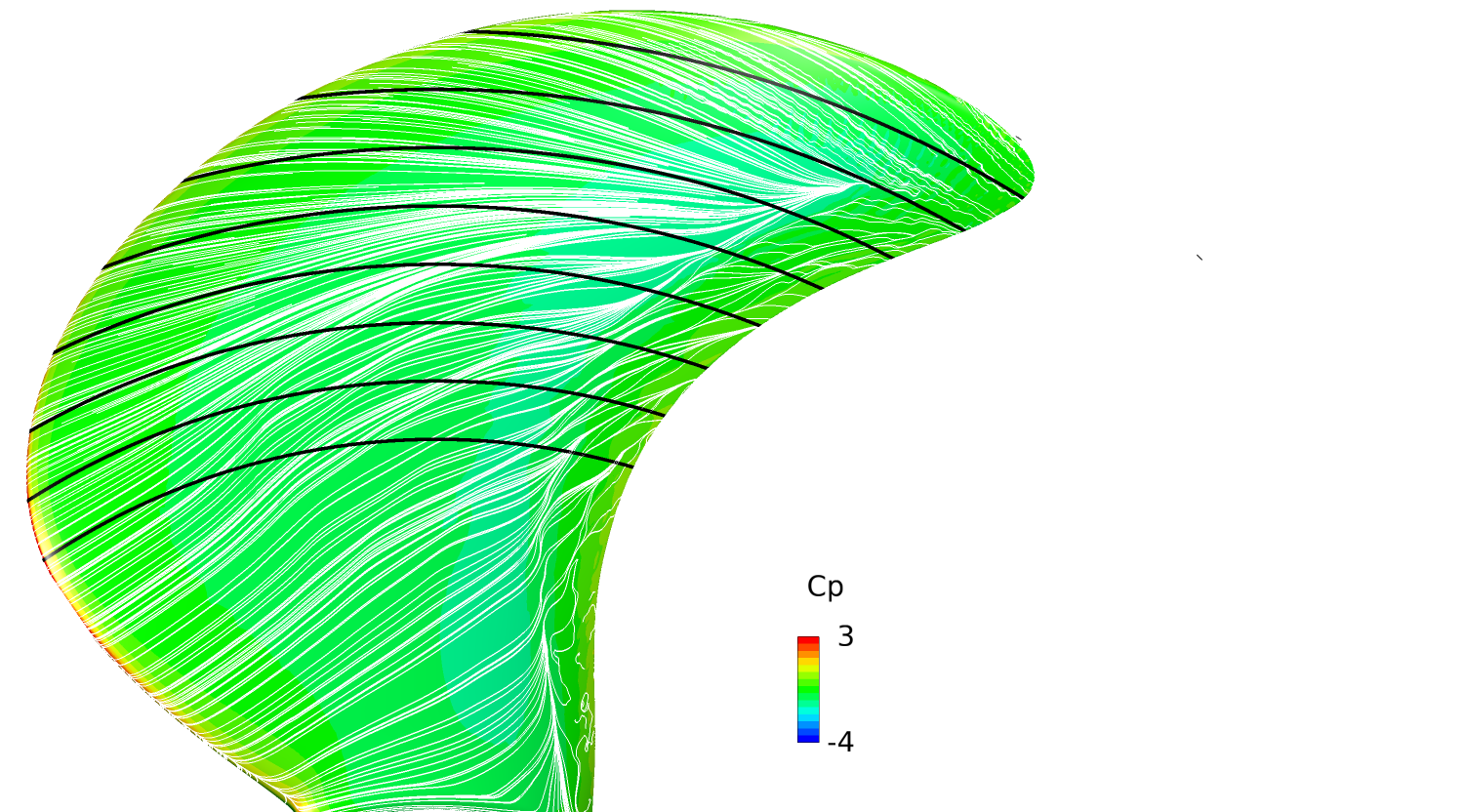}
				\put (60,60) {\makebox(0,0)[br]{\textcolor{black}{J=1.111}}}			
			\end{overpic}	
			\label{fig::smoothPropellerFPStreamlines2SS}
            \caption{Back side}						   
        \end{subfigure}
        \quad        
        \begin{subfigure}[b]{0.95\textwidth}			
			\begin{overpic}[width=0.3\textwidth]{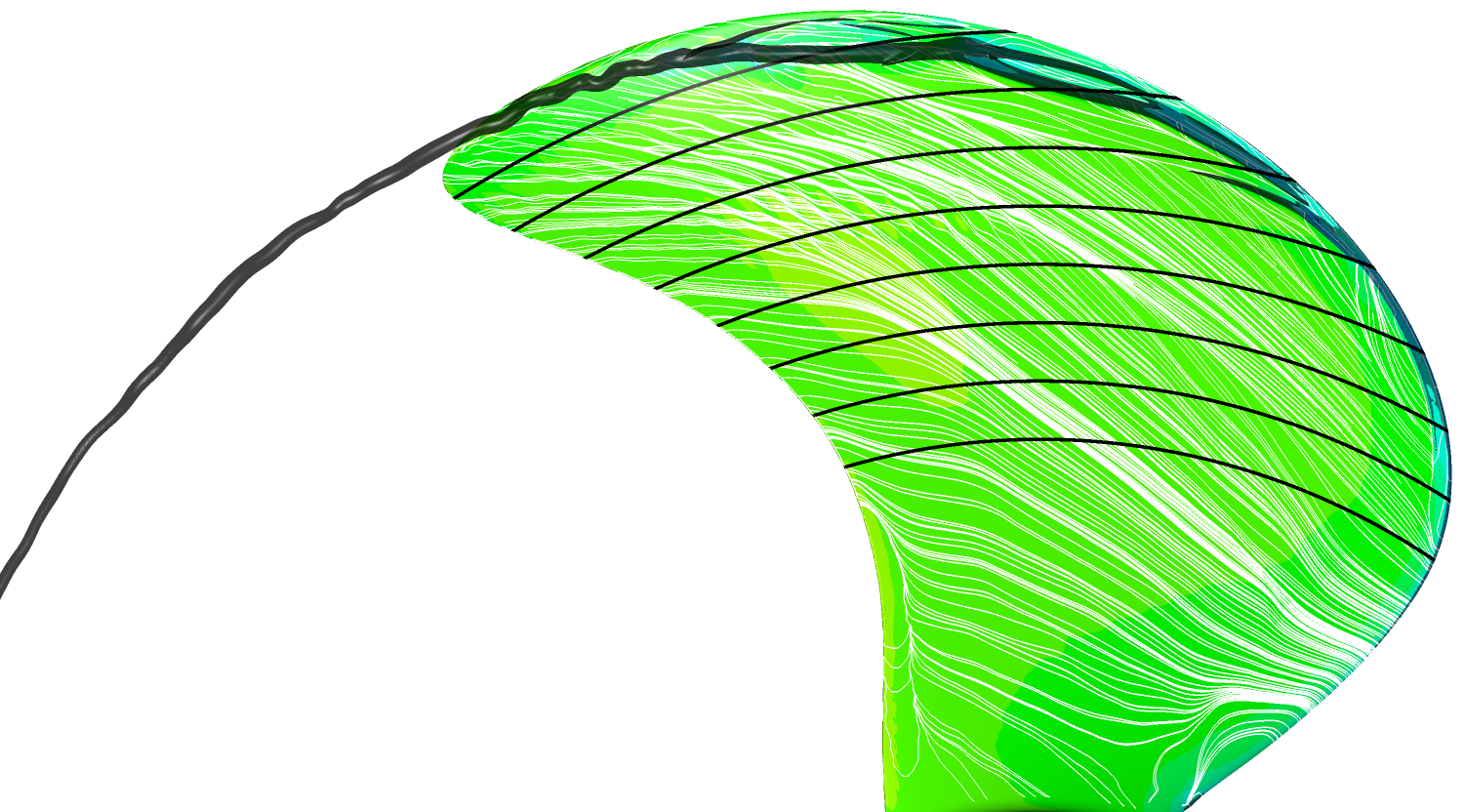}
				\put (45,16) {\makebox(0,0)[br]{\textcolor{black}{r/R=0.7}}}	
				\put(42,24.8){\vector(2,1){10}}					 
			\end{overpic}
			\begin{overpic}[width=0.31\textwidth]{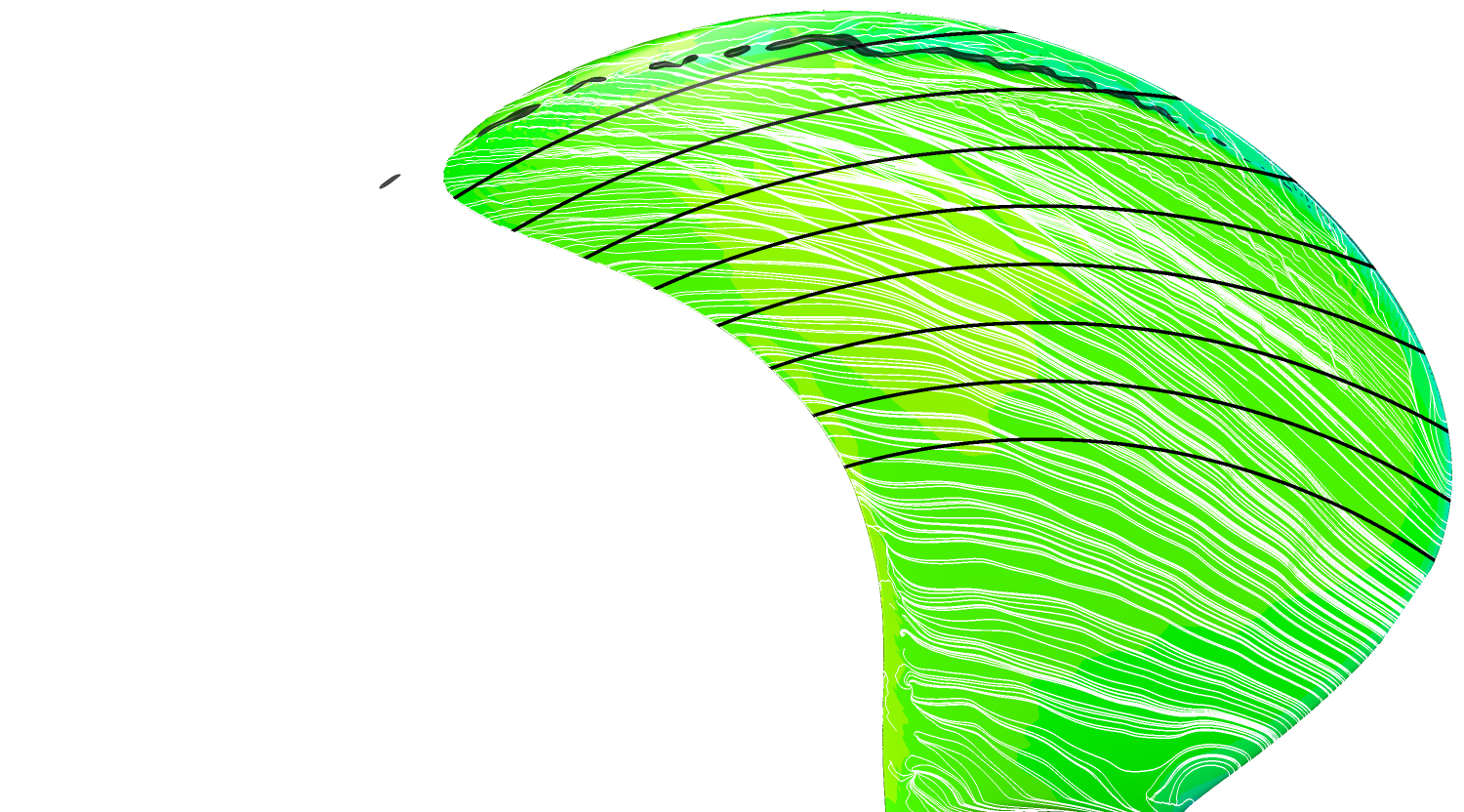} 
			\end{overpic}
			\begin{overpic}[width=0.31\textwidth]{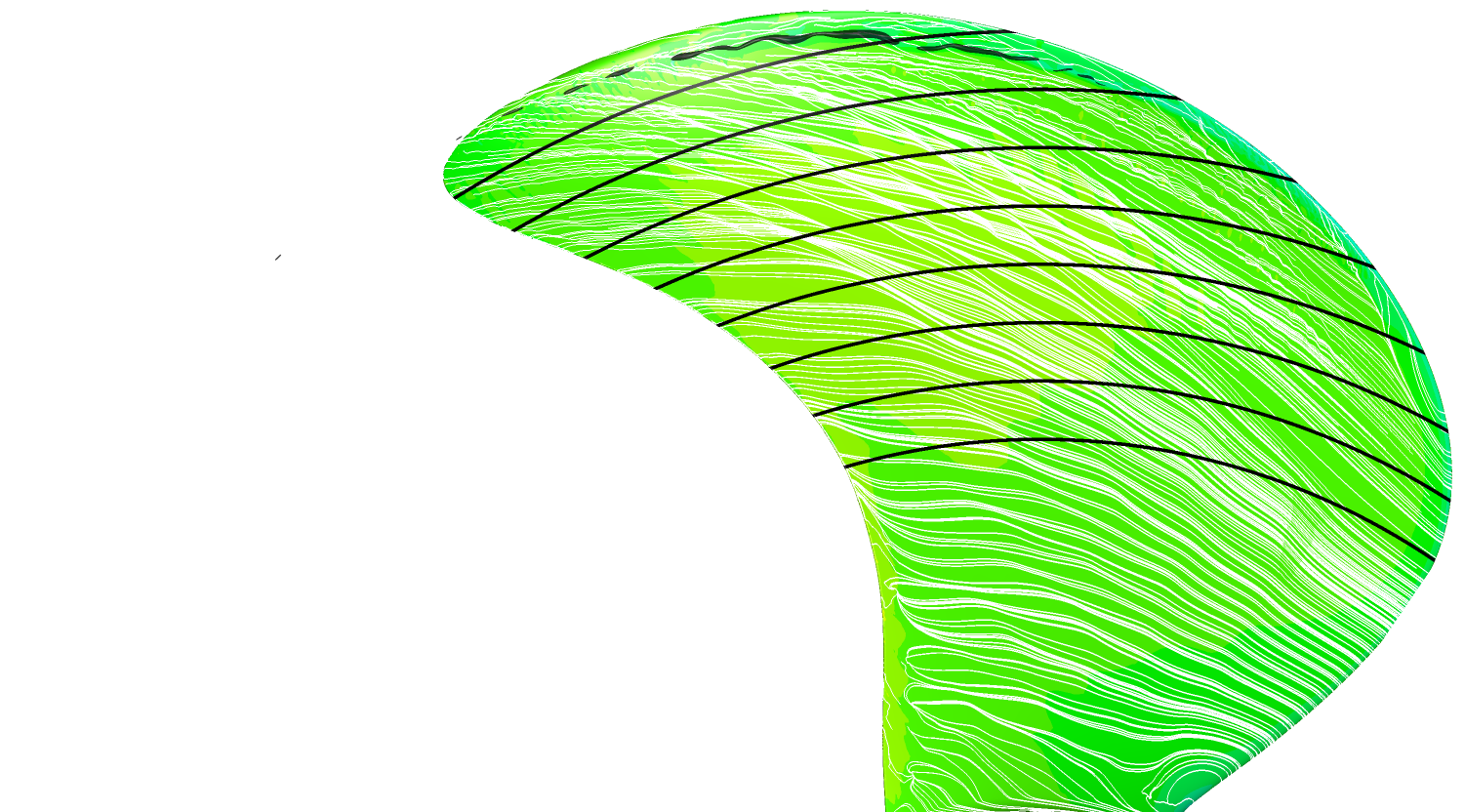}
			\end{overpic}
			\label{fig::smoothPropellerFPStreamlines2PS}			
			 \caption{Front side}							    
        \end{subfigure}                          
             \caption{Flow properties at different front side tip vortex conditions. The blade surface is coloured by $\text{C}_\text{p}$ distribution, the flow streamlines are presented in white, the tip vortex is presented by the pressure iso-surface equal to $\text{C}_\text{p}=-2$ coloured in black.} 
		\label{fig::smoothPropellerFPStreamlines2}
\end{figure}

\noindent 
The flow properties at higher J values where the tip vortex forms as a leading edge vortex are presented in Figure \ref{fig::smoothPropellerFPStreamlines2}. As expected, by increasing J values, the vortex becomes stronger and incepts at lower radii, e.g. r/R=0.7 at J=1.26. This leads to more flow suction into the rotating vortex region which can be noted from more concentrated flow streamlines on the front side in r/R$<$0.6. On the blade back side, the flow streamlines indicate a clear separation line distinguished by concentrated streamlines close to the blade trailing edge. The streamlines also highlight how the flow from lower radii moves towards the tip.

\begin{figure}[h!]
        \centering  
        \vspace{1cm} 
        \begin{subfigure}[b]{0.95\textwidth}			
			\begin{overpic}[width=0.3\textwidth]{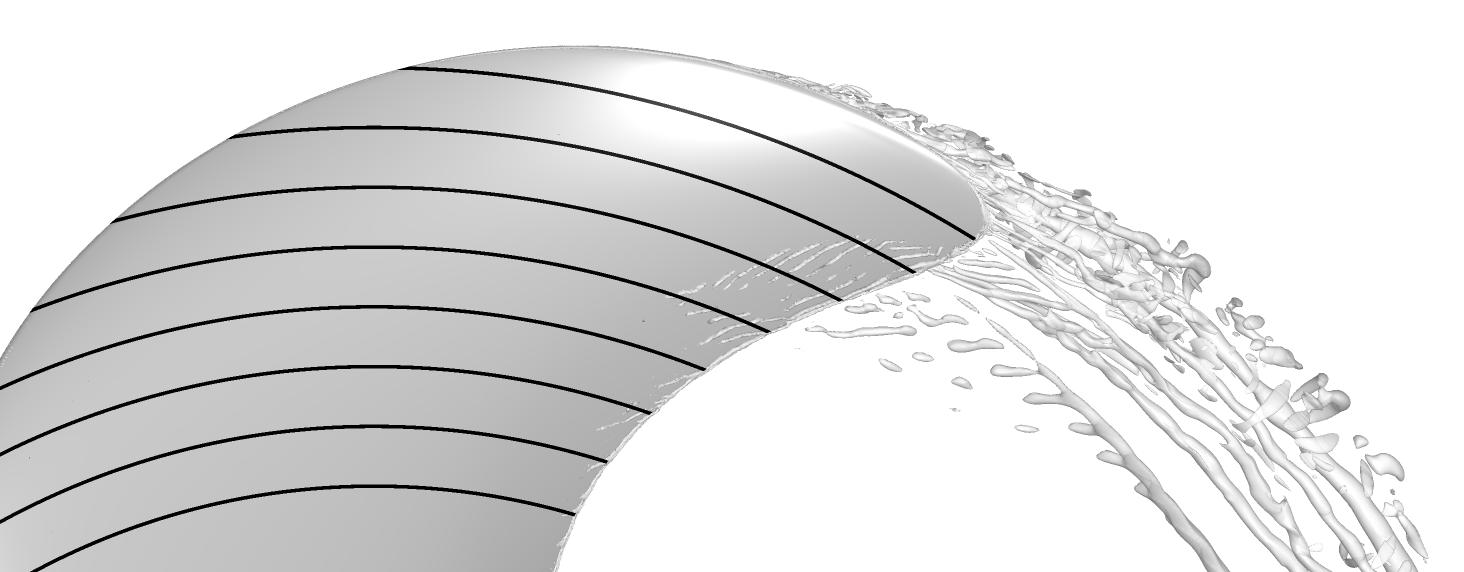} 
				\put (60,40) {\makebox(0,0)[br]{\textcolor{black}{J=1.260}}}
				\put (78,-4) {\makebox(0,0)[br]{\textcolor{black}{r/R=0.7}}}	
				\put(55.2,5.8){\vector(-2,1){10}}		
			\end{overpic}
			\begin{overpic}[width=0.31\textwidth]{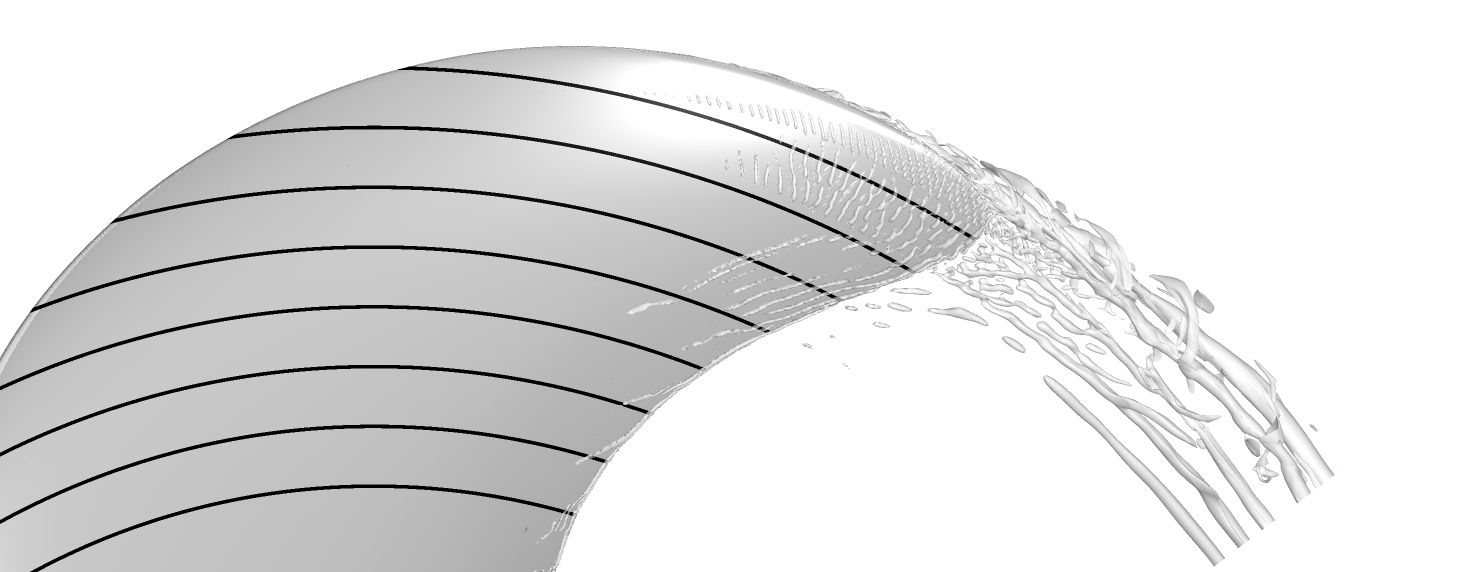} 
				\put (60,40) {\makebox(0,0)[br]{\textcolor{black}{J=1.152}}}
			\end{overpic}
			\begin{overpic}[width=0.31\textwidth]{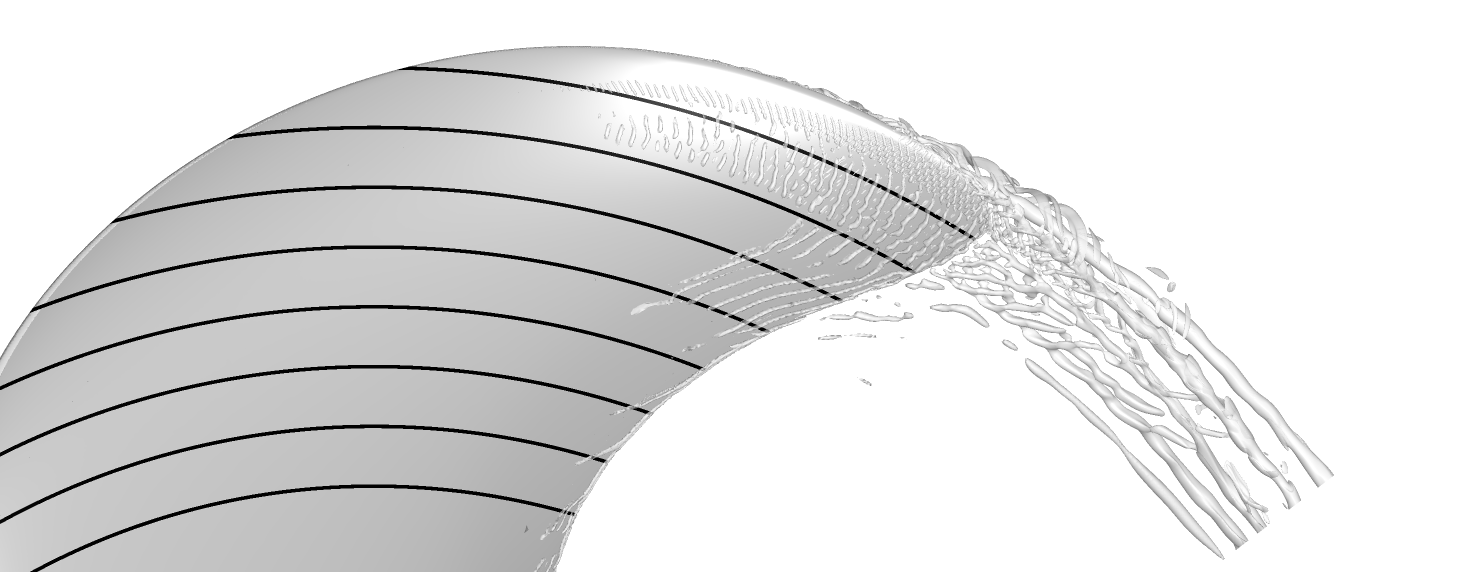} 
				\put (60,40) {\makebox(0,0)[br]{\textcolor{black}{J=1.111}}}
			\end{overpic}
			\label{fig::smoothPropellerFPQ2SS}										    
            \caption{Zoomed-view suction side}
        \end{subfigure}   

        \begin{subfigure}[b]{0.95\textwidth}			
			\begin{overpic}[width=0.3\textwidth]{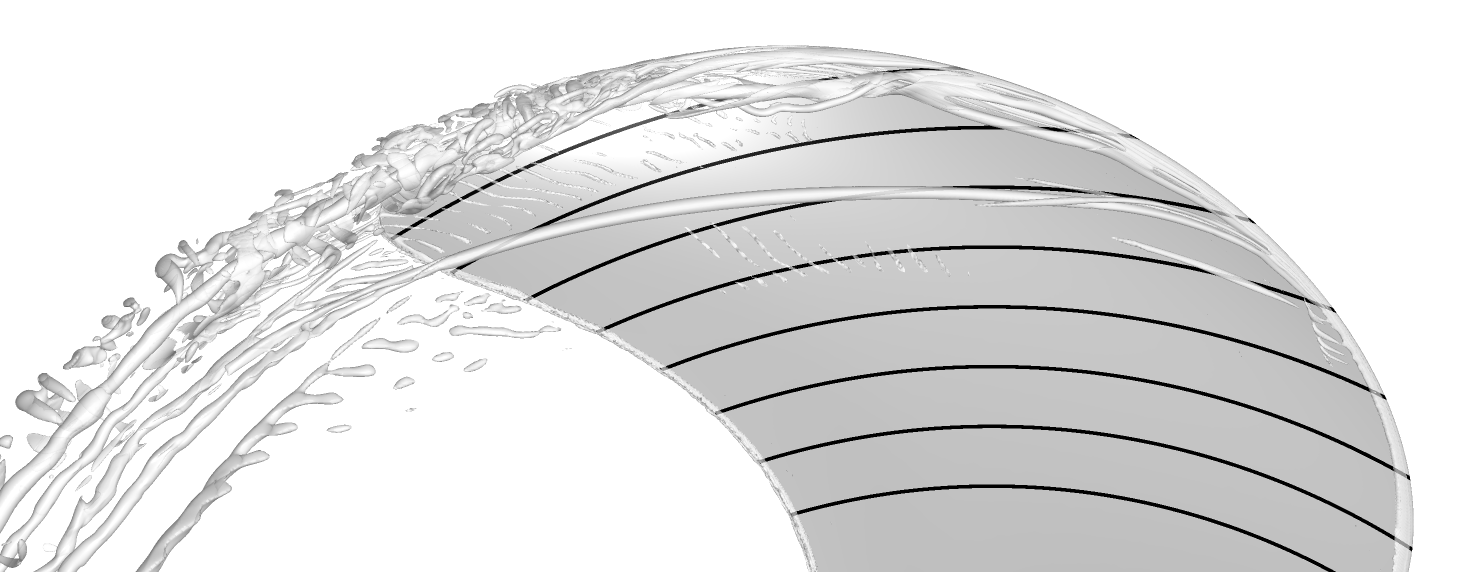} 
				\put (50,-4) {\makebox(0,0)[br]{\textcolor{black}{r/R=0.7}}}	
				\put(38,5.6){\vector(2,1){10}}				
			\end{overpic}
			\begin{overpic}[width=0.31\textwidth]{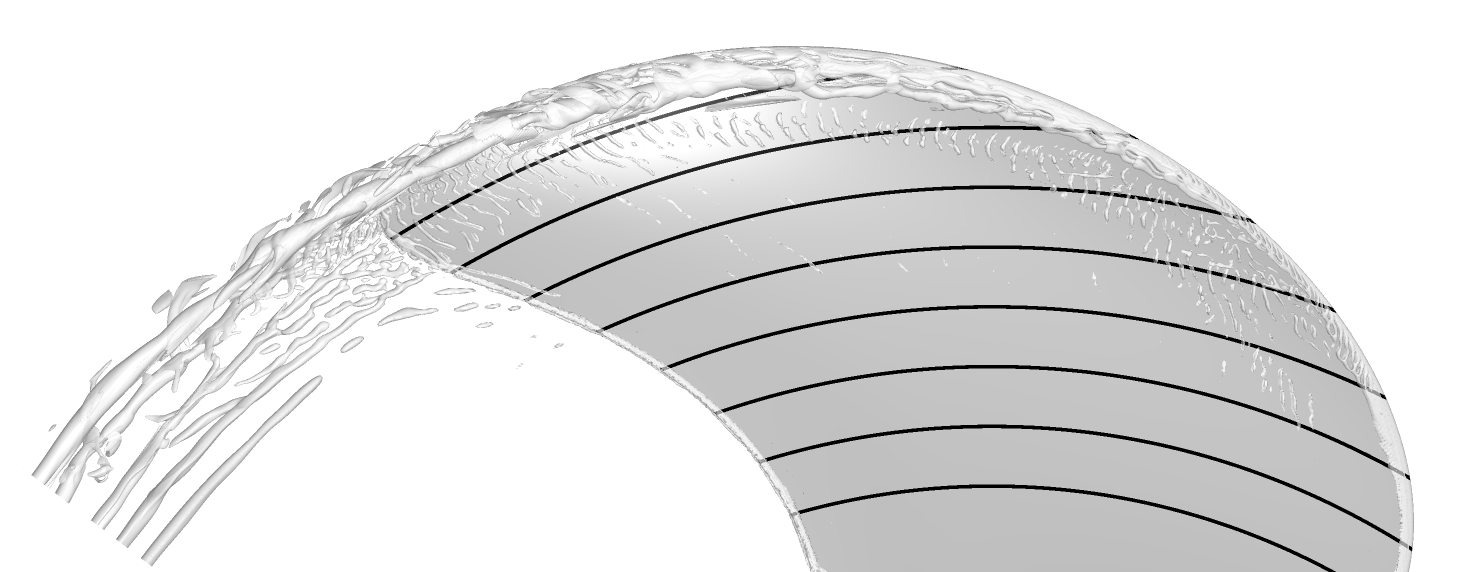} 
			\end{overpic}
			\begin{overpic}[width=0.31\textwidth]{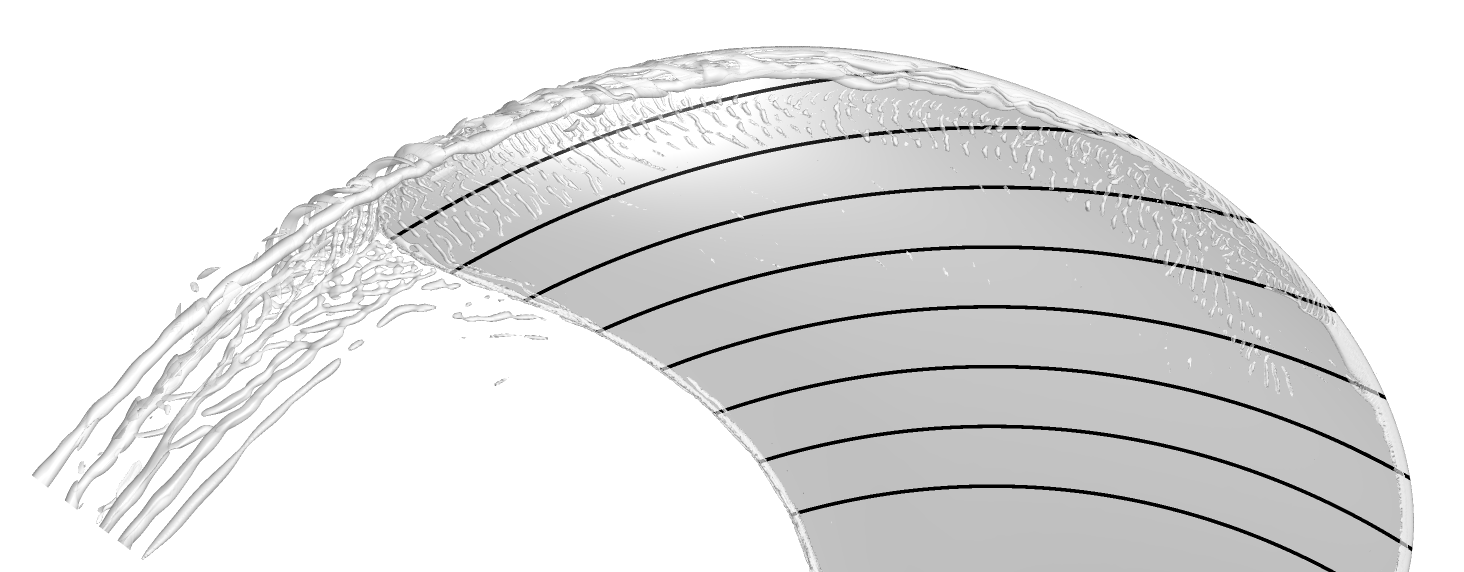} 
			\end{overpic}
			\label{fig::smoothPropellerFPQ2PS}										    
            \caption{Zoomed-view pressure side}
        \end{subfigure}  
                      
		\caption{Distribution of Q-criterion iso-surface = 200 around the blade tip at high J values. The blade surface is divided by r/R=0.05.}
		\label{fig::smoothPropellerFPQ2}
\end{figure}

\noindent 
Distribution of vortical structures presented in Figure \ref{fig::smoothPropellerFPQ2} shows how the leading edge vortex is formed.
It is clear that the trailing vortices shed from both back side and front side contribute to the tip vortex structures especially after leaving the blade. However, it is not clear whether this contribution affects the leading edge vortex strength in the region it incepts, i.e. 0.8$<$r/R$<$0.95 noted from the pressure iso-surface presented in Figure \ref{fig::smoothPropellerFPStreamlines2}. 

\noindent 
These results clarify the significance of flow structures formed in 0.7$<$r/R on the vortex strength and its development. Therefore, different roughness arrangements are considered for the front side tip vortex mitigation based on radial distances and pressure distribution, Figure \ref{fig::FrontSideroughnessSpecifications}. The summary of these arrangements and their brief descriptions are presented in Table \ref{fig::tableroughnessPatternSummary}.

\subsection{Mitigation of back side TVC}   
\noindent
Performance of the propeller for different surface roughness conditions are presented in Table \ref{table:thrustTorqueVariations} where the roughness is modelled via the rough wall function. Thrust and torque coefficients as well as the efficiency are presented relative to the smooth propeller condition at J=0.82 where the tip vortex is formed on the back side. In all of the tested roughness arrangements, the results indicate an increase in the torque coefficient when roughness is included. The thrust coefficient, however, is more dependent on the roughness pattern. For the FR blade, the maximum thrust decrease, -13.4 $\%$, and efficiency drop, -16.6 $\%$, are observed. Having roughness on the FS tip leads to higher $K_t$ but it also requires a higher $K_q$. This eventually results in a lower propeller efficiency, around -2.5 $\%$. When roughness is only applied on the BS tip, the variation of the thrust and torque is smallest. A true quantitative justification of these results demands uncertainty analysis, but it is anticipated that the trends are correctly capture at this grid refinement level. The results, however, clearly confirm that in order to minimise the negative effects of roughness on the propeller performance, the roughness area should be optimised. 
\begin{table}[h!]
\centering
\caption{Variation of thrust, torque, efficiency and TVC inception relative to the smooth propeller condition for different roughness patterns. BS: back side, FS: front side, FR: fully rough.}
\begin{tabular}{ccccc}
\hline
                   Case         				&  $K_t$ (\%) 	& $K_q$ (\%) 	& Efficiency (\%) & $\sigma_i$ (\%) \\ \hline
                   Smooth 		& -- 			& -- 			& -- 	  & -- \\   
                   BS tip 			& -0.8			& 0.2 			& -1.0	 & -16.7  \\                                                   
                   FS tip							& 1.2 			& 3.8 			& -2.5 	& -8.5   \\   
                   BS+FS tip						& 2.1 			& 4.6 			& -2.4 	& -18.2   \\     
                   FR 							& -13.4 		& 3.8 			& -16.6 	& -22.9   \\   \hline           
\end{tabular}
\label{table:thrustTorqueVariations}
\end{table}

\noindent
In Figure \ref{fig:varCavInceptRoughArea}, the predicted cavitation inception based on the minimum pressure criterion is presented for different roughness patterns defined for the back side TVC at J=0.82. As the propeller was not tested at this operating condition, the experimental data is extrapolated to this condition. As expected, the FR condition has the lowest cavitation inception compared to other roughness patterns. The predicted cavitation inception in the BS tip and BS+FS tip patterns is close to each other, and the difference between them is believed to lie in the uncertainty of the numerical results in the current simulations. 

\noindent
Even though the TVC mitigation of BS+FS tip pattern is higher, the BS tip pattern is selected to be the outcome of the roughness area optimisation for the back side TVC as it has a much lower performance degradation .

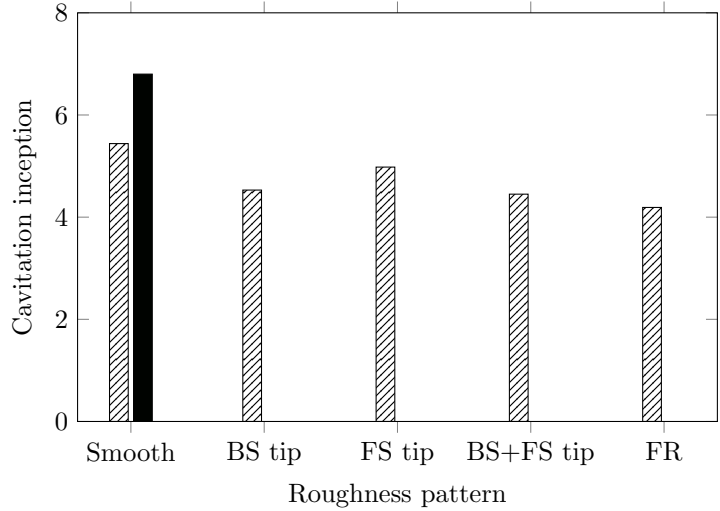
\begin{figure}[h!]
	\centering  
    \begin{tikzpicture}
        \begin{axis}[
            symbolic x coords={Smooth, BS tip, FS tip, BS+FS tip,FR},
            xtick=data,	ylabel={Cavitation inception}, xlabel={Roughness pattern}, ymin=0, ymax=8, ybar,bar width = {.7em}, width=10 cm, height=7 cm]			          
            \addplot[pattern=north east lines]  coordinates {
                (Smooth,       	5.44)
                (BS tip,   			4.53)
                (FS tip,   			4.98)
                (BS+FS tip,   		4.45)
                (FR, 			    4.19)                  
            };		            
            \addplot[fill=black!99!green] coordinates {
                (Smooth,       	6.8)  
            };
        \end{axis}
    \end{tikzpicture}
    
    \caption{Variation of the cavitation inception versus different surface roughness areas, solid bar is the extrapolated experimental measurements for the smooth blade. BS: back side, FS: front side, FR: fully rough.}
    \label{fig:varCavInceptRoughArea}    
\end{figure}

\noindent
Modelling of roughness with a wall function has some limitations, especially for the employed wall function where the roughness pattern is included into the CFD with only two representing values, i.e. roughness height and $C_s$. Including the topology of roughness elements into the computational domain is another alternative where by resolving the flow field around them more flow physics can be captured. These, however, demands for a finer computational resolution around the roughness elements and also an accurate tomography of the roughened area. As the main objective here is to discuss how results of the roughness wall function approach related to the resolving flow field approach, an arbitary roughness tomography is employed that satisfies the same averaged roughness elements height used in the wall function approach, i.e. K$_s$=250 $\mu$m.

\begin{figure}[h!]
\centering
\includegraphics[width=11cm]{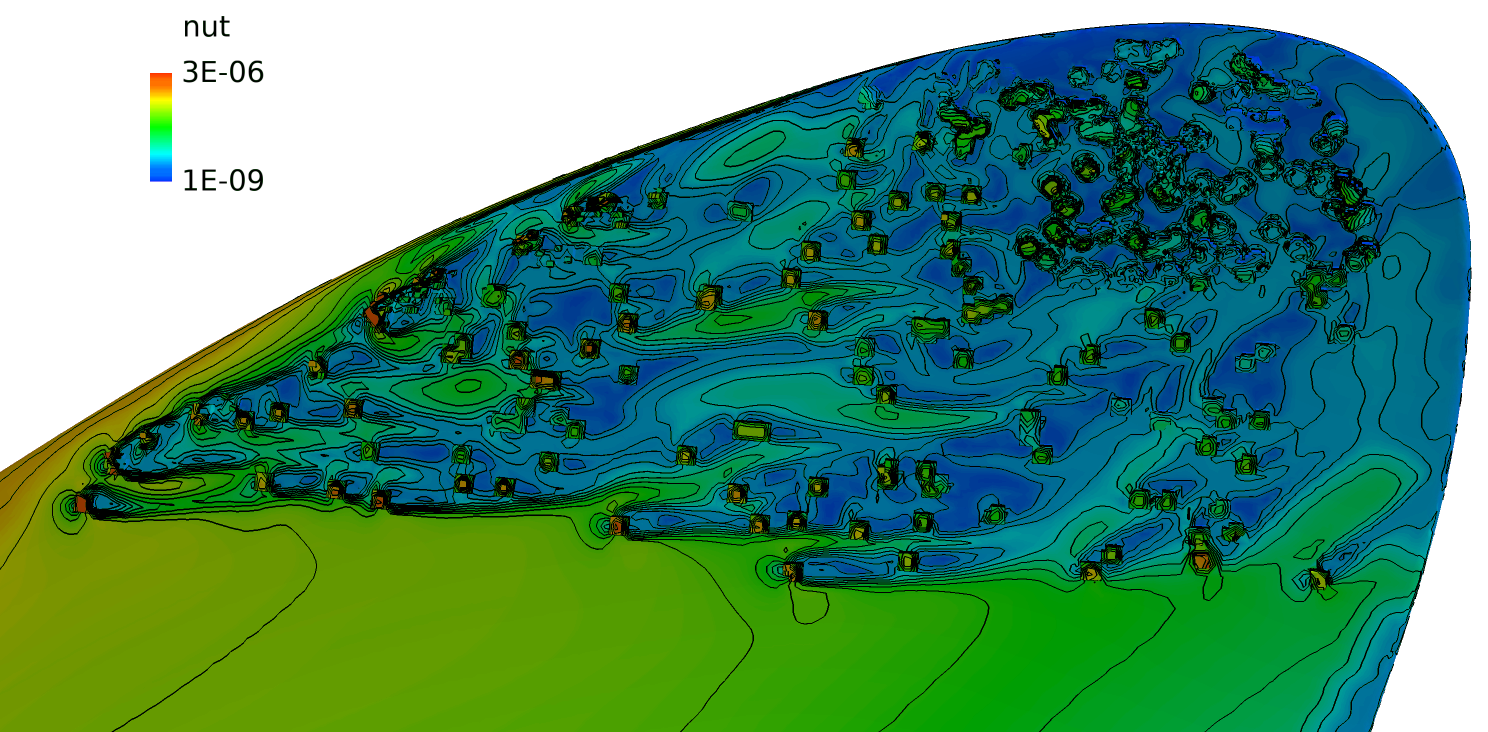}
		\caption{Distribution of the turbulent viscosity around the roughness elements, zoomed view of the roughness elements for the BS tip roughness pattern.}
		\label{fig::rough250SSRoughnesselementsNutZoomed}
\end{figure}

\begin{figure}[h!]
\centering
\includegraphics[width=14cm]{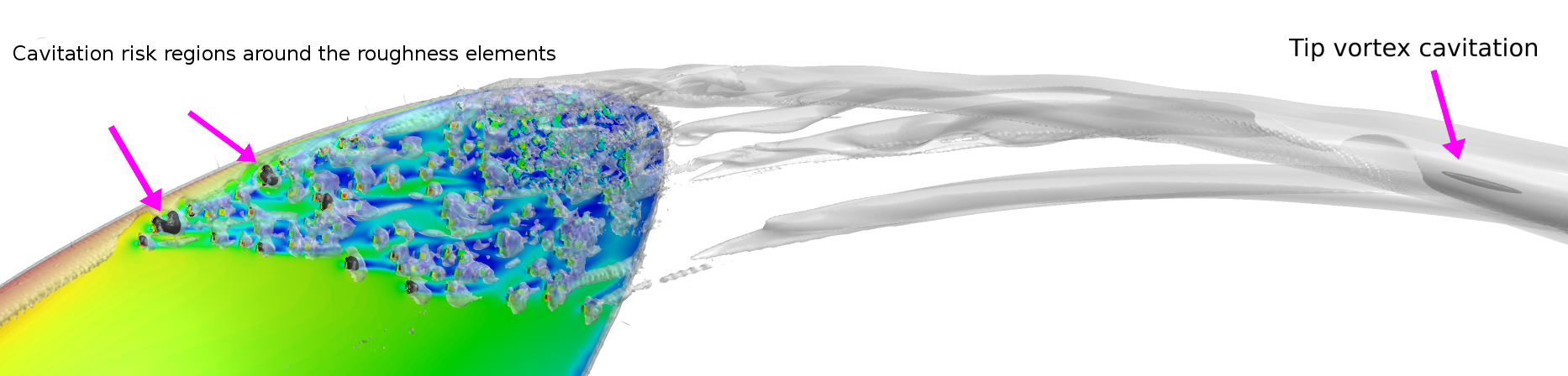}
		\caption{Distribution of the vortical structures around the roughness elements along with the iso-surface of pressure colored black, zoomed view of the roughness elements for the BS tip roughness pattern. }
		\label{fig::rough250SSRoughnesselementsQZoomed}
\end{figure}

\noindent
In Figure \ref{fig::rough250SSRoughnesselementsNutZoomed} and \ref{fig::rough250SSRoughnesselementsQZoomed}, numerical results of resolved flow around the roughness elements are presented. The figures are the zoomed view of the blade having the roughness elements on its suction side tip, i.e. the BS tip pattern. The figures include the pressure iso-surface of the saturation pressure colored black, and the vortical structure based on Q=1000 presented with transparent gray color. The low turbulent viscosity around the roughness elements indicates formation of vortical structures around them. The location of these structures are predicted by the curvature correction model, and then the turbulent viscosity is lowered there to allow the flow development. 

\noindent
Vortical structures distribution presented in Figure \ref{fig::rough250SSRoughnesselementsQZoomed} shows how flow structures generated by roughness elements interact with each other and the main tip vortices. The figure also includes the pressure iso-surface presented in colored black. The results indicate that depending on the topology of the roughness elements, their location and the propeller working conditions, roughness can increase the risk of bubble or sheet cavitation on the blade. 

\noindent
For the resolved flow around the roughness elements, the cavitation inception is found to be around 3.28 while with wall-modelling approach the predicted inception point is 4.53. Lower propeller performance is noted for the resolved flow as well, Table \ref{table:thrustTorqueVariationsSSCom}. When conducting a comparative analysis, e.g. comparing different patterns, the large difference between the cavitation inception predictions has less importance. But when it comes to find the balance between the cavitation tip vortex and the blade cavitation, the accurate prediction of flow around roughness elements is inevitable.  
\begin{table}[h!]
\centering
\caption{Variation of thrust, torque, efficiency and cavitation inception relative to the smooth foil condition for the BS tip roughness pattern with different modelling approaches.}
\begin{tabular}{ccccc}
\hline
                   Case         				&  $K_t$ (\%) 	& $K_q$ (\%) 	& Efficiency (\%)  & $\sigma_i$ (\%)\\ \hline
                   Smooth 						& -- 			& -- 			& -- 	& -- 	   \\   
                   Wall modelled ($y^+=35$)		& -0.8			& 0.2 			& -1.0	& -16.7    \\      
                   Roughness resolved ($y^+=5$)	& -1.9			& 0.1 			& -1.8	& -38.7    \\   \hline           
\end{tabular}
\label{table:thrustTorqueVariationsSSCom}
\end{table}

\subsection{Mitigation of front side TVC}  
\noindent
The roughness area optimization of the front side tip vortex is conducted at J=1.26 where more contributions of flow from lower blade radii on the tip vortex formation are observed. The flow structures on the back side are noted to affect the tip vortex properties especially downstream the tip where the trailing vortices interact with the tip vortex. However, it is not clear whether this interaction would affect the tip vortex cavitation inception on the front side of the blade. Therefore, the analysis is started by comparing having roughness all over the front side or back side of the blade, Figure \ref{fig:varCavInceptRoughArea}. The results of smooth condition as the reference condition and the fully rough condition as the maximum expected TVC mitigation condition are included.  

\noindent
Very little improvement in TVC mitigation is observed when the roughness is applied on BS of the blade compared to the smooth condition, while the results of FS and FR conditions are found to be similar. This clearly indicates that in order to mitigate the front side TVC, roughness should be applied on the front side. This agrees with our findings from the back side TVC mitigation where it is noted roughness should be applied on the same side of the blade where TVC forms.

\noindent
In order to find on which radial distance roughness should be applied to effectively mitigate TVC, different radial patterns are considered, e.g. R7080, R8090, R90100 and R70100. These patterns are illustrated in Figure \ref{fig::FrontSideroughnessSpecifications} and described in Table \ref{fig::tableroughnessPatternSummary}. The TVC inception of R7080 pattern is found to be close to the smooth condition results noting very little impact of having roughness on this area on TVC mitigation. Among the tested radial patterns the lowest TVC inception belongs to R70100 where the predicted inception is very close to the results of FS and FR patterns. This indicates the necessity of having roughness in 0.8$<$r/R$<$1.0 region in order to practically suppress front side TVC.
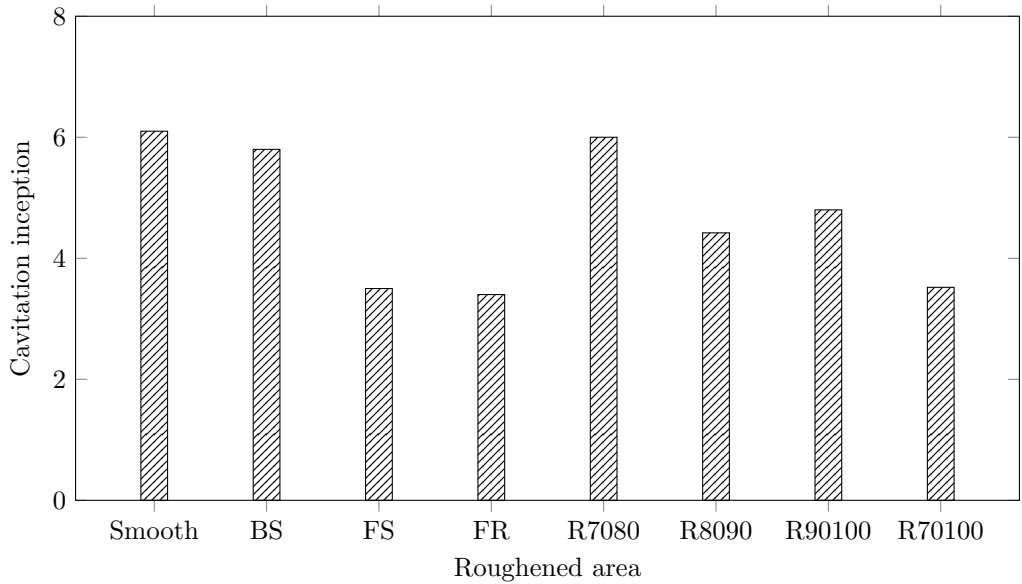
\begin{figure}[h!]
	\centering  
    \begin{tikzpicture}
        \begin{axis}[
            symbolic x coords={Smooth, BS, FS, FR, R7080, R8090, R90100, R70100},
            xtick=data,	ylabel={Cavitation inception}, xlabel={Roughened area},
             ymin=0, ymax=8, ybar,bar width = {1em},  width=14 cm, height=8 cm]	
            \addplot[pattern=north east lines]  coordinates {
                (Smooth,       	6.1)
                (BS,   			5.8)
                (FS,   			3.5)
                (FR,   			3.4)                  
                (R7080,			6.0)                  
                (R8090, 		4.42)                  
                (R90100,		4.8)                  
                (R70100,		3.52)                  
            };		            

        \end{axis}
    \end{tikzpicture}
    
    \caption{Variation of the cavitation inception versus different surface roughness areas at J=1.26, FR: fully rough foil, BS: back side roughness, FS: front side roughness, R: radial position of the roughened area on the front side of the blade.}
    \label{fig:varCavInceptRoughArea}    
\end{figure}

\noindent
Different criteria are tested to narrow down the effective areas on the blade front side in 0.8$<$r/R$<$1.0 region. Among the tested criterion, the pressure coefficient is found to be the most effective one. Based on the pressure coefficient distribution close to the blade, different roughness patterns are created, Figure \ref{fig::RE80100Distribution}. The performance of these patterns relative to the smooth condition is presented in Table \ref{table:thrustTorqueVariationsJ126Analysis}. The results show in all of the patterns both K$_T$ and K$_Q$ increase compared to the smooth condition. This has led to higher efficiency in these patterns as well. Even though the variation of efficiency is relatively small and can be assumed to lie in the uncertainty of the numerical results, it indicates that none of the patterns would have a negative impact on the propeller performance. As a result, the pattern that has the highest TVC mitigation is selected as the optimum roughness pattern for the front side TVC mitigation, i.e. RE80100. 
\begin{table}[h!]
\centering
\caption{Variation of thrust, torque, efficiency and TVC inception relative to the smooth foil condition for different roughness patterns on the front side leading edge at J=1.26, $y^+=35$.}
\begin{tabular}{ccccc}
\hline
                   Case         &  $K_t$ (\%) 	& $K_q$ (\%) 	& Efficiency (\%)   & $\sigma_i$ (\%) \\ \hline
                   Smooth 		& -- 			& -- 			& -- 	  			& -- \\   
                   RE8090 		& 2.21 			& 1.55 			& 0.64 	   			& -22.2\\   
                   RE90100 		& 4.17 			& 3.37 			& 0.78 	   			& -20.3\\     
                   RE8595 		& 3.22 			& 2.44 			& 0.75				& -29.1\\      
                   RE80100 		& 4.8 			& 3.87 			& 0.90 				& -35.2\\   \hline           
\end{tabular}
\label{table:thrustTorqueVariationsJ126Analysis}
\end{table} 

\subsection{Optimised roughness pattern}  
\begin{figure}[h!]
    \centering
					\begin{tikzpicture}
					\pgfplotsset{xmin=0.8, xmax=1.3,width=0.8\textwidth, legend pos= north east, legend columns=3}
						\begin{axis}[axis y line*=right,ymin=0, ymax=1.0,  xlabel=J,  ylabel=$\eta_0$]
						
							\addplot[only marks,mark=diamond*] table[only marks, x index=0,y index=3] 	{PLOT/1301B-OW/kT_Exp.dat};  
							\label{Exp-eff-1301B}
							\addlegendentry{Exp-$\eta_0$-MS}

							\addplot[smooth,orange] table[x index=0,y index=3] 	{PLOT/1301B-OW/MS_kT_Num.dat}; 
							\label{Num-eff-1301B}
							\addlegendentry{Num-$\eta_0$-Smooth}

							\addplot[dashed,orange] table[x index=0,y index=3] 	{PLOT/1301B-OW/MS_kT_Num_Roughness.dat}; 
							\label{Num-eff-1301BRough}
							\addlegendentry{Num-$\eta_0$-Rough}							
						\end{axis}
					
						\begin{axis}[axis y line*=left,  axis x line=none,  xlabel=J,  ymin=0, ymax=0.8,  ylabel=K$_T$ and 10K$_Q$]
							\addlegendimage{/pgfplots/refstyle=Exp-eff-1301B}\addlegendentry{Exp-$\eta_0$-MS}

							\pgfplotsset{every x tick label/.append style={font=\large, yshift=-3ex}}
							\pgfplotsset{every y tick label/.append style={font=\large, xshift=-1ex}}
			
							\addplot[only marks,mark=square*,black] table[only marks, x index=0,y index=1] 	{PLOT/1301B-OW/kT_Exp.dat};  
							\addlegendentry{Exp-K$_T$-MS}

							\addplot[only marks,mark=*,black] table[only marks, x index=0,y index=2] 	{PLOT/1301B-OW/kT_Exp.dat};
						 	\addlegendentry{Exp-K$_Q$-MS}

							\addlegendimage{/pgfplots/refstyle=Num-eff-1301B}\addlegendentry{Num-$\eta_0$-Smooth}
							
							\addplot[smooth,blue] table[x index=0,y index=1] 	{PLOT/1301B-OW/MS_kT_Num.dat};
							\addlegendentry{Num-K$_T$-Smooth}

							\addplot[smooth,red] table[x index=0,y index=2] 	{PLOT/1301B-OW/MS_kT_Num.dat};
							\addlegendentry{Num-K$_Q$-Smooth}																						

							\addlegendimage{/pgfplots/refstyle=Num-eff-1301BRough}\addlegendentry{Num-$\eta_0$-Rough}
							
							\addplot[dashed,blue] table[x index=0,y index=1] 	{PLOT/1301B-OW/MS_kT_Num_Roughness.dat};
							\addlegendentry{Num-K$_T$-Rough}

							\addplot[dashed,red] table[x index=0,y index=2] 	{PLOT/1301B-OW/MS_kT_Num_Roughness.dat};
							\addlegendentry{Num-K$_Q$-Rough}	
							
						\end{axis}
						
					\end{tikzpicture}

    \caption{Comparison of the open water performance of the model scale propeller in the smooth and optimised roughness area conditions, $y^+=35$.}
    \label{fig:comparOWForcesCoeffSmoothRoughness}
\end{figure}
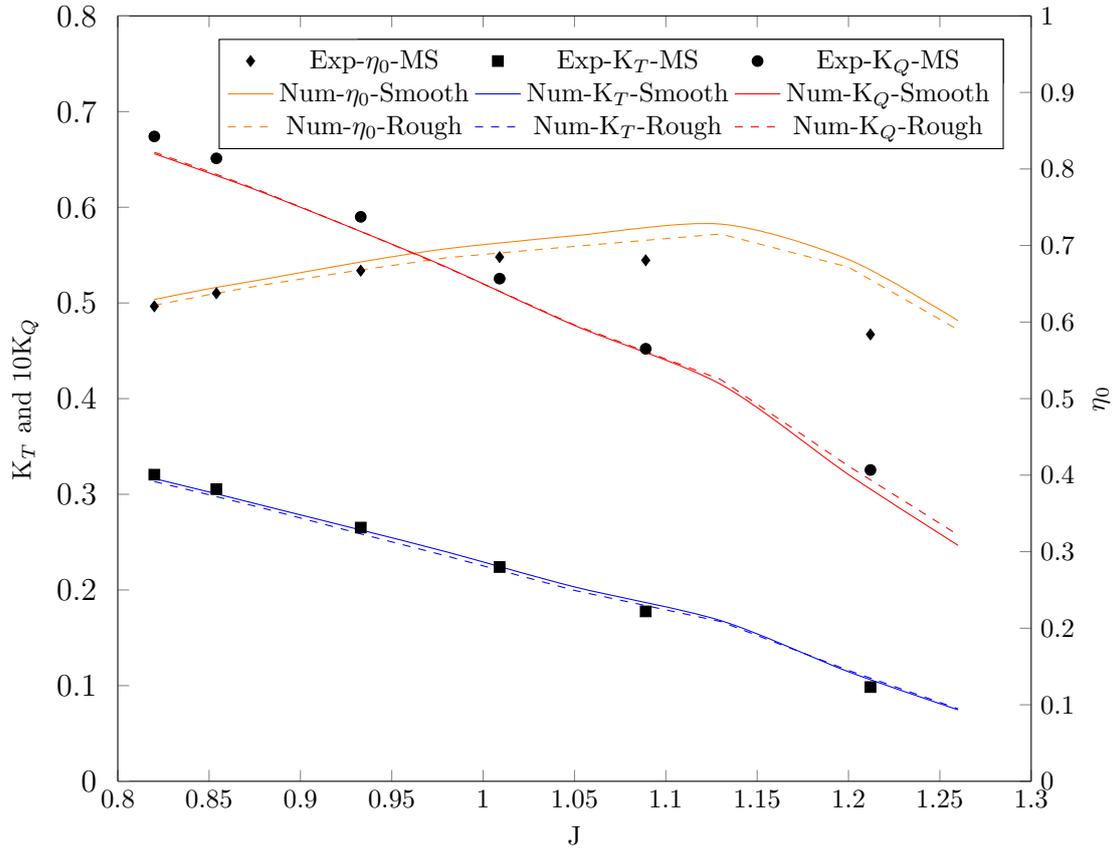

\noindent
The optimized roughness pattern is achieved by combining BS Tip pattern obtained from back side TVC mitigation study and RE80100 pattern obtained from front side TVC mitigation study. In Figure \ref{fig:comparOWForcesCoeffSmoothRoughness}, the open water performance of the model scale propeller in smooth and optimized roughness pattern (ORP) is presented. In the presented results, the roughness is modelled via the rough wall function. For J$<$1.125, similar torque coefficients are predicted in smooth and ORP while the thrust coefficient is lower in the ORP. For larger values of J, the produced thrust in smooth and ORP conditions are similar while more torque is needed in the ORP condition. This leads to having a lower efficiency in ORP condition across all of the operating conditions. Interestingly, the efficiency curves are found to be similar in smooth and ORP conditions with a small shift downward in ORP.

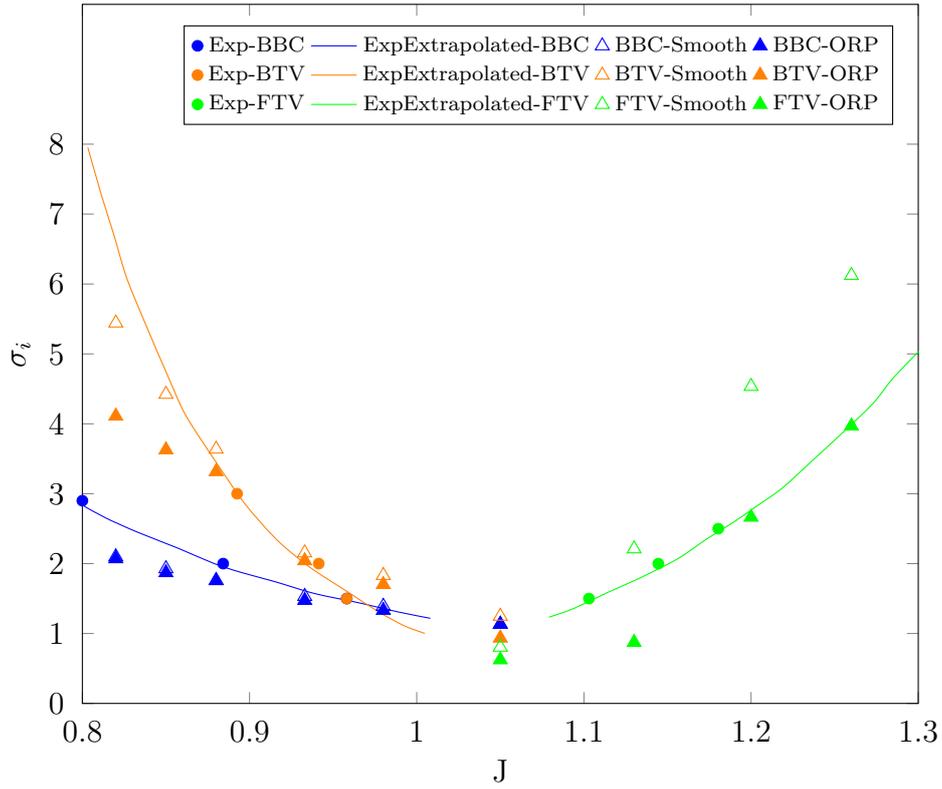
\begin{figure}[h!]
    \centering

	\begin{tikzpicture}
			\begin{axis}[font=\large,legend pos=north east, legend columns=4,  xlabel=J, ylabel=$\sigma_i$, width=0.74\textwidth, xmin=0.8, xmax=1.3, ymin=0, ymax=10, xtick={0.8,0.9,1.0,1.1,1.2,1.3}, ytick={0,1,2,3,4,5,6,7,8}]
			  
			\pgfplotsset{every x tick label/.append style={font=\large, yshift=-0.75ex}}
			\pgfplotsset{every y tick label/.append style={font=\large, xshift=-0.75ex}}
			  
			\addplot[only marks,mark=*,blue] 	table[only marks, x index=0,y index=1] 		{PLOT/1301B-OW/InceptionExp/BBC_Exp.dat};  
			\label{Exp-BBC}
			\addlegendentry{\footnotesize  Exp-BBC}

			\addplot[smooth,blue] 	table[x index=0,y index=1] 		{PLOT/1301B-OW/InceptionExp/BBC_ExpExtrapolated.dat};  
			\label{Exp-BBC}
			\addlegendentry{\footnotesize   ExpExtrapolated-BBC}

			\addplot[only marks,mark=triangle,blue, mark size=3pt] 	table[only marks, x index=0,y index=1] 		{PLOT/1301B-OW/InceptionMinPressure/BBC_minPressure.dat};  
			\label{BBC-Smooth}
			\addlegendentry{\footnotesize  BBC-Smooth}
			
			\addplot[only marks,mark=triangle*,blue, mark size=3pt] 	table[only marks, x index=0,y index=1] 		{PLOT/1301B-OW/InceptionMinPressureORA/BBC_minPressure.dat};    
			\label{BBC-ORP}
			\addlegendentry{\footnotesize  BBC-ORP}	
			
			\addplot[only marks,mark=*,orange] 	table[only marks,x index=0,y index=1] 		{PLOT/1301B-OW/InceptionExp/BTV_Exp.dat}; 
			\label{Exp-BTV}
			\addlegendentry{\footnotesize  Exp-BTV}
			
			\addplot[smooth,orange] 	table[x index=0,y index=1] 		{PLOT/1301B-OW/InceptionExp/BTV_ExpExtrapolated.dat}; 
			\label{Exp-BTV}
			\addlegendentry{\footnotesize  ExpExtrapolated-BTV}
		
			\addplot[only marks,mark=triangle,orange, mark size=3pt] 	table[only marks,x index=0,y index=1] 		{PLOT/1301B-OW/InceptionMinPressure/BTV_minPressure.dat}; 
			\label{BTV-Smooth}
			\addlegendentry{\footnotesize  BTV-Smooth}		

			\addplot[only marks,mark=triangle*,orange, mark size=3pt] 	table[only marks,x index=0,y index=1] 	{PLOT/1301B-OW/InceptionMinPressureORA/BTV_minPressure.dat}; 
			\label{BTV-ORP}
			\addlegendentry{\footnotesize  BTV-ORP}								
			
			\addplot[only marks,mark=*,green] 	table[only marks, x index=0,y index=1] 		{PLOT/1301B-OW/InceptionExp/FTV_Exp.dat};
			\label{Exp-FTC}
			\addlegendentry{\footnotesize  Exp-FTV}
	
			\addplot[smooth,green] 	table[x index=0,y index=1] 		{PLOT/1301B-OW/InceptionExp/FTV_ExpExtrapolated.dat};
			\label{Exp-FTC}
			\addlegendentry{\footnotesize  ExpExtrapolated-FTV}	

			\addplot[only marks,mark=triangle,green, mark size=3pt] 	table[only marks, x index=0,y index=1] 		{PLOT/1301B-OW/InceptionMinPressure/FTV_minPressure.dat};
			\label{FTV-Smooth}
			\addlegendentry{\footnotesize  FTV-Smooth}	

			\addplot[only marks,mark=triangle*,green, mark size=3pt] 	table[only marks, x index=0,y index=1] 			{PLOT/1301B-OW/InceptionMinPressureORA/FTV_minPressure.dat};
			\label{FTV}
			\addlegendentry{\footnotesize  FTV-ORP}	
																						
		\end{axis}
	\end{tikzpicture}

    \caption{Comparison of open water cavitation inception diagrams of the smooth and optimised roughness pattern (ORP) conditions for the model scale propeller represented by FTV, front tip vortex; BTV, back tip vortex; and BBC, back bubble cavitation, $y^+=35$.}
    \label{fig:inceptionDiagram1301BOWSmoothORP}
\end{figure}

\noindent
In Figure \ref{fig:inceptionDiagram1301BOWSmoothORP}, the cavitation inception diagram of the model scale propeller in smooth and ORP conditions are presented. The general impression is that application of roughness leads to a wider cavitation free bucket on the side, and the impact on the centre area, e.g. J=1.05, is small. This is expected as in 1.0$<$J$<$1.1 operating conditions, the dominant TVC switches from one side of the blade to the other one. This corresponds to have a weak TVC and therefore small impact of roughness on its strength. 

\noindent
More detailed comparison of TVC mitigation and performance degradation is presented in Figure \ref{fig:SSTVEffiPercentage} for the back side TVC and Figure \ref{fig:PSTVEffiPercentage} for the front side TVC. It can be noted that the average performance degradation for ORP in back side TVC mitigation is around 1.4\% and the average TVC mitigation is 14\%. The lowest impact of roughness on TVC mitigation is found to be around J=0.93.

\noindent
Compared to ORP results of the back side TVC, the impact of roughness on mitigation and performance degradation is found to be larger in the front side TVC where the average performance degradation is around 1.8\% and the average TVC mitigation is 37\%.
\begin{figure}[h!]
	\centering  
	\selectcolormodel{gray}
    \begin{subfigure}[b]{0.38\textwidth}
	\centering      	
		\begin{tikzpicture}
		    \begin{axis}[
		        symbolic x coords={0.82, 0.85, 0.88, 0.93, 0.98},
		        xtick=data, xlabel={J},	ymin=0, ymax=30, ybar, bar width = {.7em}
		      ]			          
		        \addplot[fill=black!60!green] coordinates {
		            (0.82,       24.5)
		            (0.85,       17.95)
		            (0.88,       8.59)
		            (0.93,       5.26) 
		            (0.98,       7.12) 
		        };
		    \end{axis}
		\end{tikzpicture}
	\caption{Back side TVC mitigation (\%)} 
    \end{subfigure}%
	\hspace{1.5 cm}
    \begin{subfigure}[b]{0.38\textwidth}
    	\centering  	
		\begin{tikzpicture}
		    \begin{axis}[
		        symbolic x coords={0.82, 0.85, 0.88, 0.93, 0.98},
		        xtick=data,	xlabel={J}, ymin=0, ymax=4, ybar,bar width = {.7em}
		      ]			          
		        \addplot[pattern=north west lines]  coordinates {
		            (0.82,       1.207)
		            (0.85,       1.218)
		            (0.88,       1.088)
		            (0.93,       1.536) 
		            (0.98,       1.697) 
		        };
		    \end{axis}
		\end{tikzpicture}
	\caption{$\eta_0$ degradation (\%)} 		
    \end{subfigure}%
    \caption{Percentage of the back side TVC mitigation and open water efficiency drop in different operating conditions for the optimum roughness pattern relative to the smooth propeller condition, $y^+=35$.}
    \label{fig:SSTVEffiPercentage}    
\end{figure}
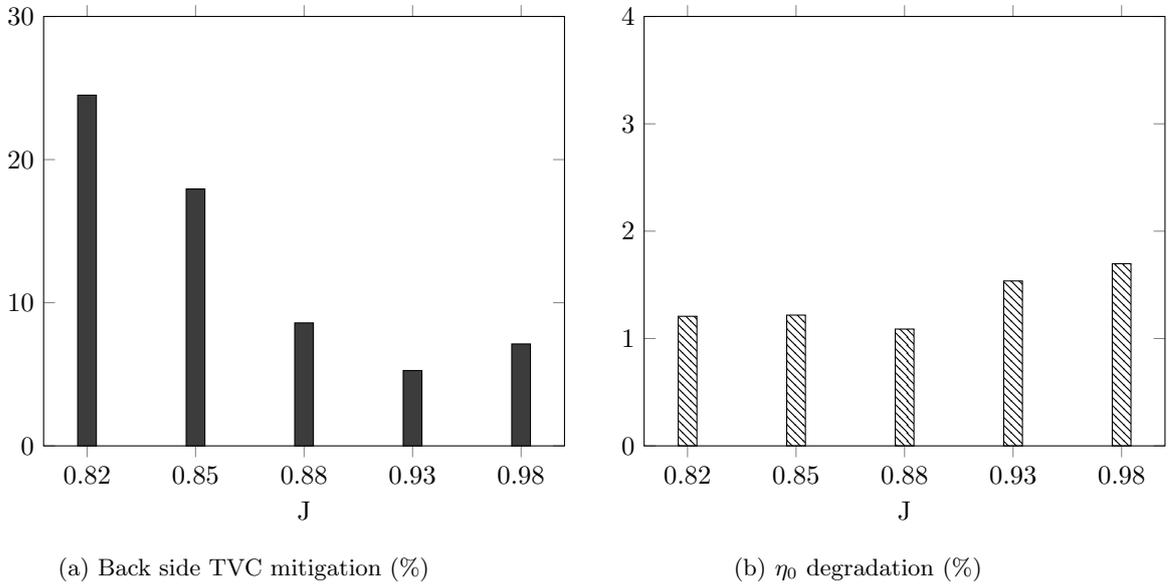 
\begin{figure}[h!]
	\centering  
	\selectcolormodel{gray}
    \begin{subfigure}[b]{0.38\textwidth}
	\centering      	
		\begin{tikzpicture}
		    \begin{axis}[
		        symbolic x coords={1.05,1.13,1.20,1.26},
		        xtick=data, xlabel={J},	ymin=0, ymax=70, ybar, bar width = {.7em}
		      ]			          
		        \addplot[fill=black!60!green] coordinates {
		            (1.05,       22.54)
		            (1.13,       60.51)
		            (1.20,       41.25)
		            (1.26,       35.19) 
		        };
		    \end{axis}
		\end{tikzpicture}
	\caption{Pressure side TVC mitigation (\%)} 
    \end{subfigure}%
	\hspace{1.5 cm}
    \begin{subfigure}[b]{0.38\textwidth}
    	\centering  	
		\begin{tikzpicture}
		    \begin{axis}[
		        symbolic x coords={1.05,1.13,1.20,1.26},
		        xtick=data,	xlabel={J}, ymin=0, ymax=4, ybar,bar width = {.7em}
		      ]			          
		        \addplot[pattern=north west lines]  coordinates {
		            (1.05,       1.95)
		            (1.13,       1.78)
		            (1.20,       1.47)
		            (1.26,       1.71) 
		        };
		    \end{axis}
		\end{tikzpicture}
	\caption{$\eta_0$ degradation (\%)} 		
    \end{subfigure}%
    \caption{Percentage of the pressure side TVC mitigation and open water efficiency drop in different operating conditions for the optimum roughness pattern, $y^+=35$.}
    \label{fig:PSTVEffiPercentage}    
\end{figure}
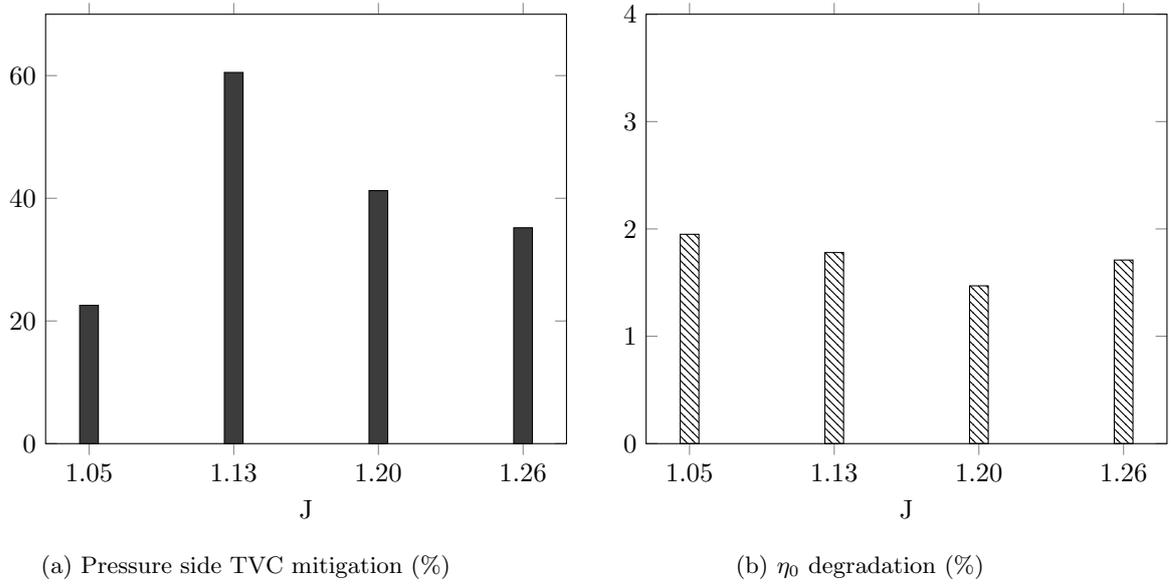

\subsection{Optimized roughness pattern in full scale condition}
\begin{figure}[h!]
	\centering  
	\selectcolormodel{gray}
    \begin{tikzpicture}
        \begin{axis}[
            symbolic x coords={J=0.82, J=0.93, J=1.26},
            xtick=data,	ylabel={TVC mitigation (\%)}, xlabel={},
             ymin=0, ymax=50, ybar,bar width = {1em},  width=7 cm, height=8 cm]	
            \addplot[fill=black!60!green]  coordinates {
                (J=0.82,       	37.6)                
                (J=0.93,      	10.4)                
                (J=1.26,       	19.3)                                
            };		            
        \end{axis}      
    \end{tikzpicture}
    \hspace{1 cm}
    \begin{tikzpicture}
        \begin{axis}[
            symbolic x coords={J=0.82, J=0.93, J=1.26},
            xtick=data,	ylabel={Performance degradation (\%)}, xlabel={},
             ymin=0, ymax=5, ybar,bar width = {1em},  width=7 cm, height=8 cm]	
            \addplot[pattern=north east lines]  coordinates {
                (J=0.82,       	1.32)                
                (J=0.93,      	1.53)                
                (J=1.26,       	1.36)                                
            };		            
        \end{axis}      
    \end{tikzpicture}    
    \caption{The optimum roughness area results for the full scale propeller. TVC and propeller performance of each operating condition is normalised by the smooth propeller results of that condition.}
    \label{fig:FSPerTVCOptimumAreaRough}    
\end{figure}
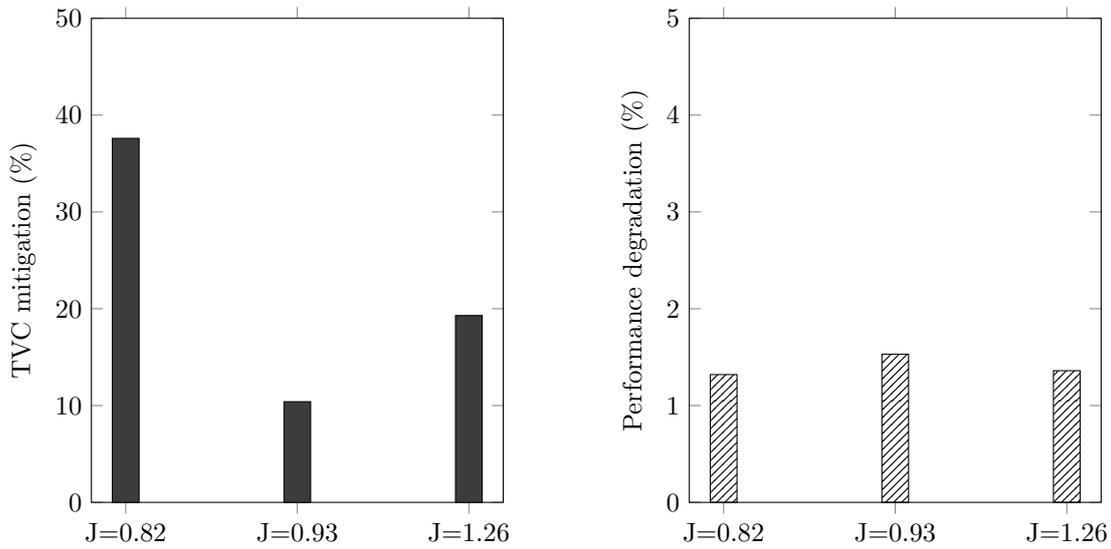

 \noindent
In Figure \ref{fig:FSPerTVCOptimumAreaRough}, TVC mitigation and performance degradation of the full scale propeller at three different operating conditions are provided. These simulations are performed on the propeller with $y^+=50$ where the roughness elements having $K_s^+=925$ are incorporated into the computational domain, Table \ref{fig::tableFSMSDetails}. Similar to the model scale results, the lowest TVC mitigation and highest performance degradation are found to be at the design point, i.e. J=0.93. The results indicate an average TVC mitigation of 22\% and performance degradation of 1.4\% by employing optimised roughness pattern.

\noindent
Resolving the flow around the roughness elements provides the possibility of investigating the risk of bubble or sheet cavitation formation at these spots. At the tested conditions, no obvious increase in bubble or sheet cavitation due to the roughness elements is observed.

\section{Conclusion}
\noindent 
The results on the tested propeller in model scale and full scale conditions show that the application of roughness on a specified and restricted area on the blade can be a solution to improve inception characteristics of a tip vortex while keeping the propeller performance degradation in a reasonable range. Roughness elements interact with the boundary layer developed on the blade and consequently alter its distribution and properties. If the area where the roughness is applied coincides with the area where the tip vortex forms, this interaction affects the tip vortex formation leading to a less concentrated tip vortex, i.e. tip vortex mitigation.

\noindent 
The negative effects of roughness on the propeller performance can be minimized when the roughness area is optimized based on the flow properties and structures effective in tip vortex formation and development. For the evaluated propeller design, two distinct types of tip vortices are observed. At lower advance ratio numbers, the vortex forms on the tip of the blade which can incept either on the tip or slightly downstream depending on the tip vortex strength dependency on the roll-up process. For this type of tip vortex, application of roughness on the blade tip region where the vortex forms is found to be effective. The other type of tip vortex appears at higher J values as leading edge tip vortex where roughness application on the limited area of leading edge on the same side of the vortex roll-up is found to be effective. The optimized pattern that can be used across different operating conditions are obtained by simultaneous application of roughness on these two areas.

\noindent 
It is noted that application of roughness leads to a wider cavitation free bucket where its impact in the operating conditions close to the design point is lower compared to the conditions with higher loads on the blade. The findings show that the optimized roughness pattern in the model scale condition leads to an average TVC mitigation of 37\% with an average performance degradation of 1.8\% while in the full scale condition an average TVC mitigation of 22\% and performance degradation of 1.4\% are obtained

\noindent
In the obtained results, no obvious indication of increasing the risk of bubble or sheet cavitation around the roughness elements is observed. This agrees well with our previous numerical and experimental investigations on the TVC mitigation around an elliptical foil.
There are, however, a few key factors in practical application of roughness for TVC mitigation that demands further considerations and investigations. The free stream turbulence variation, i.e. wake flow effect, or the roughness impact on the cavitating tip vortex conditions are two examples of these key factors. Even though, the numerical results presented in this paper clearly show the capability of roughness application in mitigation of tip vortex flows in both model scale and full scale conditions. We showed if the roughness pattern is optimized with respect to the tip vortex flow properties, a reasonable balance between the TVC mitigation and performance degradation can be achieved.   
\section{Acknowledgements}
\label{Acknowledgements}
Financial support for this work has been provided by VINNOVA through the RoughProp project, Grant number 2018-04085, and Kongsberg Maritime Sweden AB through the University Technology Centre in Computational Hydrodynamics hosted at the Department of Mechanics and Maritime Sciences at Chalmers. The simulations are performed on resources at Chalmers Centre for Computational Science and Engineering (C3SE) provided by the Swedish National Infrastructure for Computing (SNIC).

\newpage    
    \bibliographystyle{unsrt}
    \bibliography{main}   
        
\end{document}